\documentclass[useAMS,usenatbib,usegraphicx]{mn2e}
\usepackage{times,amssymb}

\usepackage[numbers]{natbib}
\bibpunct{(}{)}{;}{a}{}{,}

\begin{document}

\title[Galaxy Disk in Equilibrium Models]{Stellar and Gaseous Disk Structures in Cosmological Galaxy Equilibrium Models}
\author[B. Rathaus and A. Sternberg]{Ben Rathaus, Amiel Sternberg \\
Raymond and Beverly Sackler, School of Physics and Astronomy, \\
Tel-Aviv University, Ramat-Aviv, 69978, Israel}

\date{E-mail: ben.rathaus@gmail.com\\ E-mail: amiel@astro.tau.ac.il \\ \\ In original form: \today{}} 
\pagerange{\pageref{firstpage}--\pageref{lastpage}}
\maketitle\label{firstpage}
\begin{abstract}
We present ``radially-resolved-equilibrium-models" for the growth of stellar and gaseous disks in cosmologically accreting massive halos.  Our focus is on objects that evolve to redshifts $z\sim 2$.  We solve the time-dependent equations that govern the radially dependent star-formation rates, inflows and outflows from and to the inter- and circum-galactic medium, and inward radial gas flows within the disks.  The stellar and gaseous disks reach equilibrium configurations on dynamical time scales much shorter than variations in the cosmological dark matter halo growth and baryonic accretions rates.  We show analytically that mass and global angular momentum conservation naturally give rise to exponential gas and stellar disks over many radial length scales. As expected, the gaseous disks are more extended as set by the condition Toomre $Q<1$ for star-formation.  The disks rapidly become baryon dominated.  For massive, $5\times 10^{12}M_\odot$ halos at redshift $z=2$, we reproduced the typical observed star-formation rates of $\sim 100 \, M_\odot \, {\rm yr}^{-1}$, stellar masses $\sim 9\times 10^{10}\, M_\odot$, gas contents $\sim 10^{11}\, M_\odot$, half mass sizes of 4.5 and 5.8 kpc for the stars and gas, and characteristic surface densities of $500$ and $ 400\, M_\odot \, {\rm pc}^{-2}$ for the stars and gas.
\end{abstract}


\section{Introduction}
Given the diversity of mass and length scales, redshifts, initial conditions and physical processes that take part in the hierarchical formation of galaxies, the fact that most galaxies exhibit some common key properties, such as an exponential disk profile e.g. \citet{expNew2,expNew, expO-1,expO-2}, and follow common scaling relations such as star formation rate-stellar mass relation (e.g. \citealt{scaling-5,scaling-3,scaling-2,scaling-4,Dave-1,scaling, Whitaker}) is remarkable. Such simple relations may imply that despite the diversity of initial and end states, there is a ``subset" of universal processes that effectively dominate these relations. Even though many constituents are necessary to model correctly galaxy formation to its last detail, a toy model that involves as few parameters and processes as possible, and can nevertheless yield those key properties, is powerful in its ability to offer intuition towards their understanding.

This approach has motivated the \emph{`equilibrium-model'} (or `bathtub-model') \citep{Bouche,Dave-2,Lilly,Dekel} that has been used to investigate the content of galaxies, both gaseous and stellar, the associated  star formation rate (SFR), metallicities, and the relations between them, especially for the (blue) ``main sequence" of galaxies. These models typically assume an equilibrium relation (see for example \citealt{Dave-2}), $\dot{M}_* = M_{\rm{gas}}/t_{\rm{dep}}$, where $\dot{M}_*$ is the SFR, $M_{\rm{gas}}$ is the gas mass of the disk and $t_{\rm{dep}}$  is the gas-to-star depletion time (or ``turn-over" time, \citet{Larson}). In addition, the outflowing baryonic mass out of the galaxy, back to the circum- or intergalactic medium (CGM/IGM) in these models is proportional to the SFR, $\dot{M}_{\rm{out}}=\eta \dot{M}_*$, where $\eta$ is the mass-loading factor, set by energy and momentum injection back into the interstellar medium (ISM).

To date, all of the equilibrium models have been applied to galaxies as a whole, ignoring the internal disk structures. In contrast, in this paper we formulate a radially resolved model, in which the equilibrium relations are ``local" and functions of (planar) distance from the disk centres, and assuming that the gas dynamics is set by the mass-continuity equation and global angular momentum conservation. Our model is self-consistent, as star formation, disk rotation, outflows and the disk size are all set dynamically with the disk evolution. We study in detail how systems that follow these simple assumptions evolve, reach a (quasi-)steady state, accumulate stellar mass, and analyze dynamically their surface density and Toomre-$Q$ profiles. In our presentation, we show that many of our numerical results can be derived analytically, by solving simplified cases of the full (partial differential) equations, for our single phase ISM.

The paper is organised as follows. In \S\ref{sec:setup} we write the evolution equations and the assumptions of our model, and discuss a dimensionless representation for it. In \S\ref{sec:cosSnap} we consider a cosmological snapshot -- in which we hold the cosmological conditions constant, and analyze the consequent steady state and the convergence to it. Next, in \S\ref{sec:CosEvol} we generalise the cosmological snapshot, by accounting for the evolution of the halos, and varying baryon accretion rates. In \S\ref{sec:results} we analyze the results of our model for the cosmologically evolving system, and present a parameter study. We summarise in \S\ref{sec:discussion}.

\section{Equilibrium Disk Model}
\label{sec:setup}
\subsection{Physical ingredients}
We consider an evolving, rotationally supported, star forming disk, within a cosmologically growing DM halo. The halo accretes baryonic matter from its cosmological neighbourhood, at some given rate $\dot{\Sigma}_{\rm{in}}$, forms stars continuously, and expels matter back to the CGM/IGM at a rate proportional to the star formation rate. Our radially resolved equilibrium model is based on the dynamics set by mass-continuity and angular momentum conservation, together with local equilibrium relations.

The mass continuity equation in cylindrical coordinates (assuming cylindrical symmetry) includes cosmological accretion, and star formation and outflows as source and sink terms, and follows the gas surface density in the disk, $\Sigma(t,r)$. It can therefore be written as
\begin{eqnarray}
\frac{\partial}{\partial t}\Sigma + \frac{1}{r} \frac{\partial}{\partial r} (r v_r \Sigma) &=& \dot{\Sigma}_{\rm{in}} - \dot{\Sigma}_* - \dot{\Sigma}_{\rm{out}} \nonumber\\
&=& \dot{\Sigma}_{\rm{in}} - \frac{1+\eta}{t_{\rm{dep}}}\Sigma ~~~,\label{eq:dimen}
\end{eqnarray}
where $ \dot{\Sigma}_*$ and $\dot{\Sigma}_{\rm{out}}$ are the ``local" SFR and outflow rates respectively, $v_r= v_r(t,r)$ is the radial gas velocity within the disk (inflow velocity is $<0$), and where for the second equality we have assumed `local' equilibrium relations, 
\begin{eqnarray}
\dot{\Sigma}_*(t,r) &=& \Sigma(t,r)/t_{\rm{dep}} \nonumber\\
\dot{\Sigma}_{\rm{out}}(t,r) &=& \eta\dot{\Sigma}_*(t,r)~~~.\label{eq:EQ}
\end{eqnarray}
To solve the mass-continuity equation, we will have to specify the boundary condition at some outer radius $R$, and the gas radial inflow velocity $v_r$ at each radius. We do that as follows:
\begin{itemize}
\item We assume that the entire (thin) baryonic disk, gas+stars, is confined to   a radius $R_d$, so that $\Sigma(r>R_d)=0$.
We determine $R_d$ by setting the specific angular momentum of the accreted gas onto the disk, at any time $t$, equal to that of the parent halo with a given (cosmological) spin parameter
\begin{equation}
\lambda \equiv \frac{J|E|^{1/2}}{G M^{5/2}} ~~~,
\end{equation}
where $J,E$ and $M$ are the angular momentum, energy and the virial mass of the halo, respectively, and $G$ is Newton's gravitational constant. 
Observations and simulations suggest that the specific angular momentum of the baryonic disks indeed remain globally equal to the parent halos as the disks evolve despite the complex inflow and outflow patterns (e.g. \citealt{Danovich,GenelNew,Burkert2}). We therefore do not impose cosmological initial conditions for the cumulative mass distribution of specific angular momentum found in dark matter only simulations \citep[e.g.][]{Bullock}. 

\item In realistic disks, the radial gas inflows depend in a complex way on local conditions and global tidal torques and bar instabilities. In our model we assume an azimuthally averaged gas inflow velocity, $v_r$,  at each radius $r$, and we consider three cases. The first two are (\textit{a}) constant radial inflow velocity, $v_r=\rm{const.}$, and (\textit{b}) radial inflow velocity that satisfies $v_r = v_R\sqrt{R_d/r}$, where $v_R$ is the inflow velocity at the outer disk radius $R_d$. The third case is motivated by a viscous disk in which the dissipated energy in one rotational timescale, which is of order $ c_s^2\Omega $, where $\Omega=v_{\rm rot}/r$ and $v_{\rm{rot}}$ is the rotational velocity at each radius, is compensated by the gain in potential energy, $-v_{\rm rot}^2 r/v_r$ \citep{Gammie, Cacciato}, leading to (\textit{c}) $v_r=-c_s^2/v_{\rm{rot}}$.
Our case (\textit{c}) in combination with our mass conservation Eqn.~(\ref{eq:dimen}) is similar to the viscous evolution models presented by \citet{viscous1} and \cite{viscous2} but with the inclusion of inflow and outflows, and self consistent star formation.
\end{itemize}

With the gas surface density $\Sigma$ we can keep track of the gas mass in the disk out to distance $r\leq R_d(t)$ as a function of time
\begin{equation}\label{eq:mass}
M_{\rm{gas}}(t,r) = 2\pi \int_{0}^{r}{{\rm{d}}r' r' \Sigma(t,r') } ~~~.
\end{equation}
The total gas mass in the disk is $M_{\rm{gas}}(t,R_d(t))$. To obtain the accumulated stellar mass $M_*(t,r)$ one has to integrate the SFR over time, and assume some stellar migration profile. We will assume in this paper that once formed, stars do not migrate radially, but rotate around the disk center together with the inflowing gas, so that 
\begin{equation}
M_*(t,r) = 2\pi \int_{t_0}^{t} {\rm{d}}t'\,\int_{0}^{r}{{\rm{d}}r' r' \dot\Sigma_*(t,r') } ~~~,
\end{equation}
where $t_0$ denotes a time before star formation has begun.

Solving for $\Sigma$ will also enable us to examine the local Toomre-$Q$ parameter \citep{ToomreQ}, that quantifies the stability of the gas in the disk with respect to star formation, as a function of the galactocentric distance. By definition,
\begin{equation}\label{eq:Q}
Q(r,t)\equiv \frac{c_s \kappa}{\pi G \Sigma} ~~~,
\end{equation}
where $c_s$ is the gas sound speed (thermal plus turbulent) and $\kappa$ is the epicyclic frequency 
\begin{equation}\label{eq:kappa}
\kappa^2 = 4\Omega^2 + r\frac{\rm d}{{\rm{d}} r}\Omega^2 ~~~,
\end{equation}
We approximate $v_{\rm{rot}}$ by the Keplerian rotational velocity for a spherical mass distribution $M(r)$, that is composed of the gaseous and stellar disk components, plus the extended dark matter distribution
\begin{equation}\label{eq:vrot}
v_{\rm{rot}} = \sqrt{\frac{GM(r)}{r}} ~~~.
\end{equation}
We assume that the rotation curves for our two-dimensional disks differ only slightly than for spherical mass distributions (see \citet{BT}). We ignore pressure support along the disk plane (c.f. \citet{Burkert}).

\subsection{Dimensionless representation}
It is useful to rescale to  dimensionless form. We define  dimensionless time and distance coordinates, 
\begin{eqnarray}
\tau &\equiv& t/t_{\rm{crs}} \nonumber \\
x&\equiv& r/R_d ~~~,
\end{eqnarray}
where $t_{\rm{crs}}$ is the crossing time, 
\begin{equation}
t_{\rm{crs}}=R_d/\left|v_R\right| ~~~,
\end{equation}
The normalised inflow velocity is 
\begin{equation}\label{eq:u}
u\equiv v_r/\left|v_R\right| ~~~.
\end{equation}
We then define  the dimensionless ``accumulation" parameter 
\begin{equation}
\alpha=\frac{t_{\rm{crs}}(1+\eta)}{t_{\rm{dep}}} ~~~,
\end{equation}
and the normalised surface density
\begin{equation}
y \equiv \frac{\Sigma}{t_{\rm{crs}}\dot{\Sigma}_{\rm{in}}} ~~~.
\end{equation}
Thus in dimensionless form the mass-continuity equation is
\begin{equation}\label{eq:cont}
\frac{\partial}{\partial \tau} y +\frac{1}{x} \frac{\partial}{\partial x} (x u y) =  1 - \alpha  y ~~~.
\end{equation}
This dimensionless representation is valid only if  $\dot{\Sigma}_{\rm{in}}$  and $v_R$ are constant, which is a reasonable assumption only if the disk reaches a steady-state faster than the time it takes $\dot{\Sigma}_{\rm{in}}$ to change significantly. We assume this in \S\ref{sec:cosSnap}, but relax this assumption in \S\ref{sec:CosEvol} , which will force us back to Eqn.~(\ref{eq:dimen}).

In terms of this dimensionless notation, the fixed constant  inflow velocity is written as $u(x) = -1$, the $v_r\propto1/\sqrt{r}$ case is written as $u(x) = -1/\sqrt{x}$ and the $v_r\propto1/v_{\rm rot}$ case can be written as $u=-A\sqrt{x/m}$, where 
\begin{equation}\label{eq:m}
m(x)=\int_{0}^{x}{\rm d}x'x'y(x')
\end{equation}
is a ``dimensionless mass", and the coefficient
\begin{equation}
A\equiv \frac{c_s^2}{R_d\sqrt{2\pi v_R G \dot\Sigma_{\rm in}}}
\end{equation}
is set by the input parameters.

\section{Cosmological Snapshot}
\label{sec:cosSnap}

We start by solving the disk evolution equation for a ``cosmological snapshot", for which the halo mass and cosmological inflow rates are constant, and the dimensionless version of our model is valid. We focus first on steady state solutions with (\S\ref{sec:YSF}) and without (\S\ref{sec:NSF}) star formation. We demonstrate that simple mass conservation naturally leads to disks with exponential profiles over many length scales. Finally, in \S\ref{sec:CCA} we examine the timescale for the convergence to steady-state.

\subsection{No star formation ($\alpha=0$)}
\label{sec:NSF}
Quantitatively, the case of no star formation (i.e. infinite depletion time), is considered simply by setting $\alpha=0$ in Eqn.~(\ref{eq:cont}). In steady state, the inflow velocity is independent of time, and the continuity equation can be easily solved, yielding
\begin{equation}\label{eq:yQSS}
y(x) = \frac{x^2-1}{2x}\frac{1}{u(x)} ~~~.
\end{equation}
This solution is valid for $u=-1$ or $u = -1/\sqrt{x}$ (cases (\textit{a}) and (\textit{b})) discussed above. For $u=-A\sqrt{x/m}$ (case (\textit{c})) Eqn.~(\ref{eq:yQSS}) is also satisfied but depends on $m(x)$, so this is actually a differential equation on its own, rather than a solution for the surface density. Ignoring the DM halo, i.e., assuming that the entire rotation curve, and hence the inflow velocity are determined solely by the baryonic matter in the disk, this differential equation can also be solved analytically. Using Eqn.~(\ref{eq:m}) the differential equation for the surface density can be written as
\begin{equation}
\frac{1}{\sqrt{m}}\frac{{\rm d}m}{{\rm d}x} = 2\frac{\rm d}{{\rm{d}}x} \sqrt{m} = \frac{1}{2A}\left( x^{-1/2} - x^{3/2}  \right) ~~~.
\end{equation}
Solving this differential equation, setting the constant of integration so that $m(0) = 0$, and writing the solution in terms of $y=({\rm d}m/{\rm d}x)/x$, we get
\begin{equation}\label{eq:wow}
y(x)= \frac{x^2-1}{2x}\frac{5 \left( 1-6x^2/5 + x^4/5\right)}{4\left( x^2-1\right)} ~~~.
\end{equation}
Comparing this solution to Eqn.~(\ref{eq:yQSS}) shows that in the no-star formation no-DM case, the constant inflow velocity is a better fit to the physically motivated case (\textit{c}), in which the inflow velocity changes by $\sim20$ per cent across the entire disk.

\begin{figure}
\centering
\includegraphics[width=0.8\columnwidth]{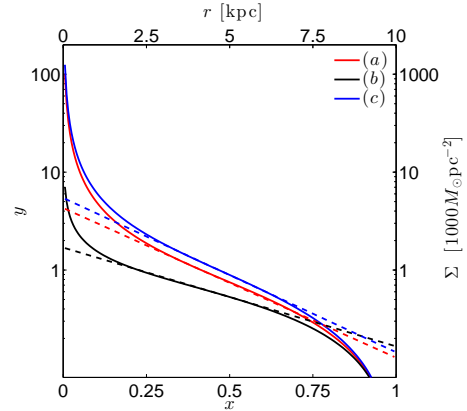}
\caption{Steady-state solution of the radially resolved equilibrium model in its simplest form -- with no star formation, and for the three simple radial gas inflow cases discussed in the text, in dimensionless representation (bottom-left axes) and in a dimensional setting (top-right axes). For all three cases an exponential surface density fit is plotted (dashed lines, same colours). It can be seen that the exponential fit works well for large portions of the disks, except near the disk centers, and at the disk outskirts. }
\label{fig:yQSS}
\end{figure}

The dimensionless gas profiles, for all three inflow models, are generic results for  $\alpha=0$, and do not depend on the accretion rate, the size of the disk or the sound speed. It is instructive to display these solutions dimensionally for a specific (characteristic) cosmological accretion rate  $\dot{\Sigma}_{\rm{in}}$ and disk inflow velocity  $v_R$. In Fig.~\ref{fig:yQSS} 
we plot the steady state solution for these three $u$-cases, in a dimensionless form (bottom-left axes), and dimensional setting (top-right axes), for $R=10\,{\rm{kpc}}$, $v_R=15\,\rm{km}\,\rm{s}^{-1}$ and $\dot{\Sigma}_{\rm{in}}\approx 1500 \,M_\odot\,\rm{pc}^{-2}\rm{Gyr}^{-1}$, a cosmological accretion rate that corresponds to a uniformly distributed cosmological accretion by a $5\times10^{12}M_\odot$ median halo at $z=2$ \citep{Dekel-2}.
\begin{figure*}
\centering
\includegraphics[width=\columnwidth]{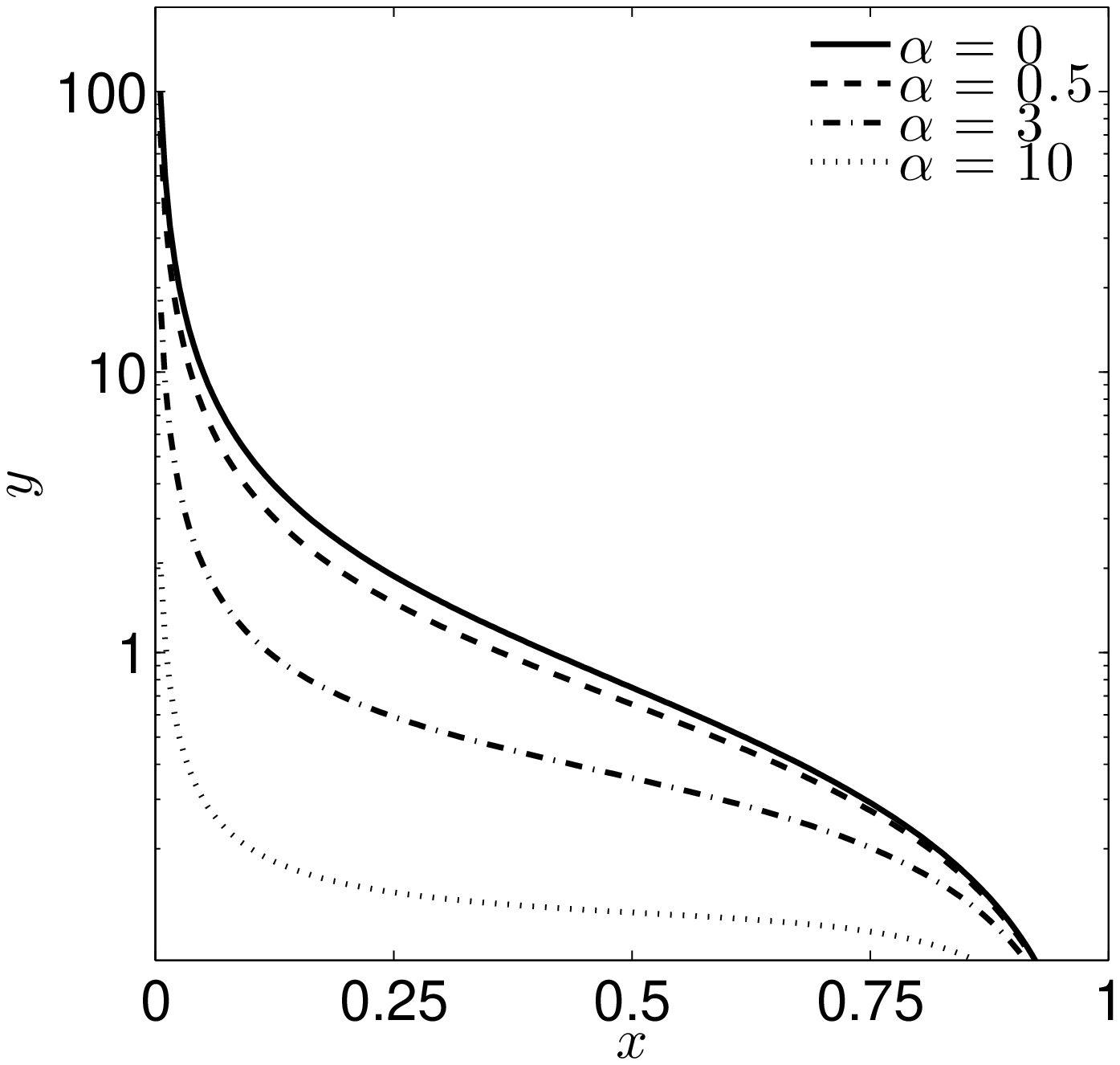}
\includegraphics[width=\columnwidth]{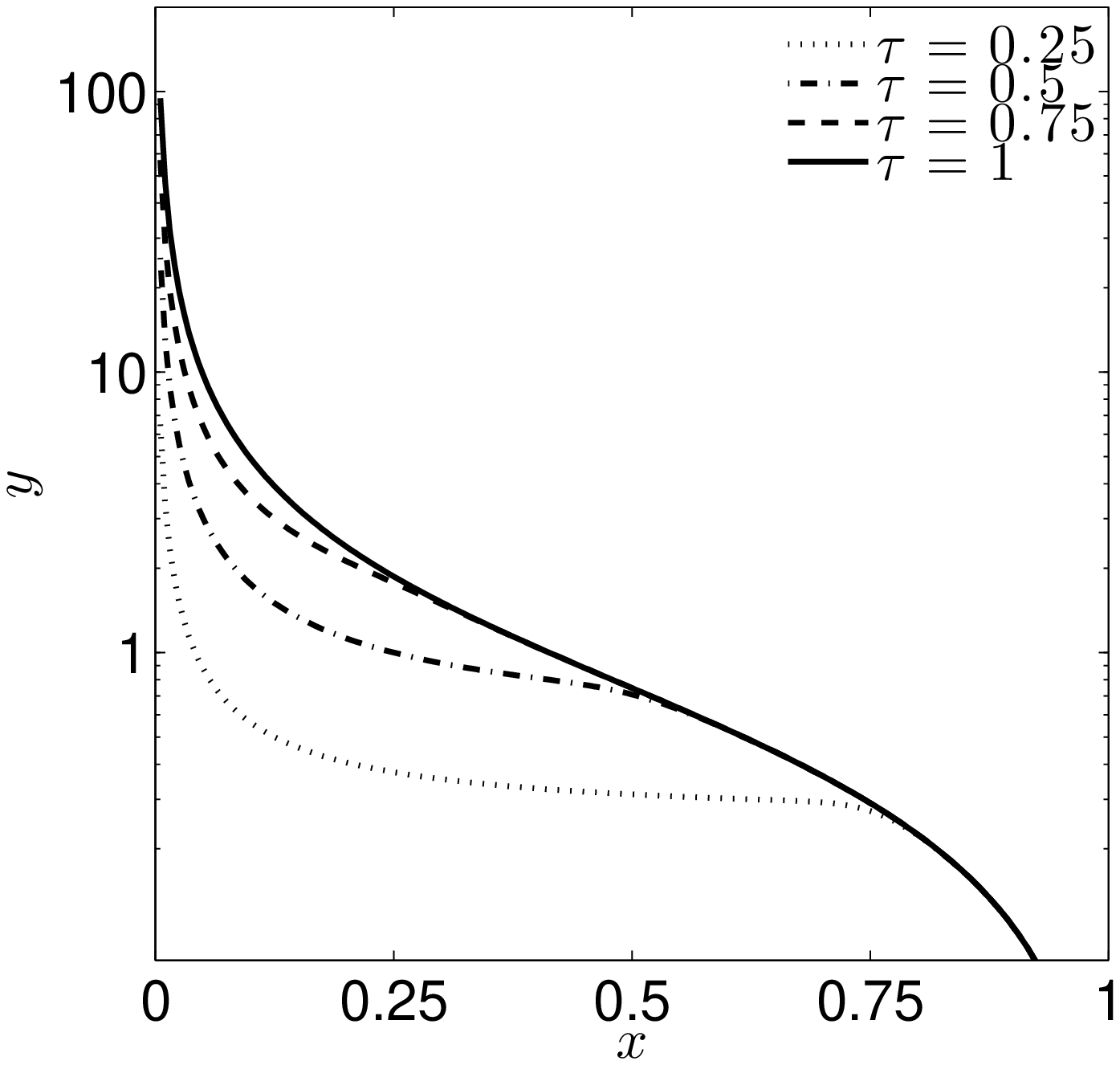}
\caption{(\emph{left}) Analytic solutions for the steady state normalised  surface densities of the gas in the disk, $y$, for a few $\alpha$ values ($\alpha=0$ means no star formation at all). It can be seen that as $\alpha$ increases the surface density has a shallower slope.
(\emph{right}) A numerical time dependent solution of the dimensionless continuity equation, Eqn~(\ref{eq:dimen}), starting with an empty disk, for no-star formation ($\alpha=0$), with $u=-1$, at a few time steps $\tau$. It can be seen that $y$  converges to its steady state solution within a crossing time $\tau\sim1$ (that is, $t_{\rm{crs}}$). We note that case (\textit{c}) takes a shorter to reach a steady state, $\tau\sim0.75$. }\label{fig:y}
\end{figure*}

It can be seen, that the surface density in all three cases, fits an exponential profile (dashed lines, same colours) quite well for the bulk of the disk. At the inner part of the disk, however, the exponential profile ceases to fit the steady state solution, and in fact, the surface density diverges towards the center of the disk. However, the enclosed mass $M_{\rm{gas}}(r)$ for these cases does not diverge.
The gas profile shapes we find and display in Figs.~\ref{fig:yQSS} and \ref{fig:y} resemble those found assuming the "strong form" of angular momentum conservation, such that the cumulative distribution of mass versus specific angular momentum, $M(<j)$, is conserved and equal to that of a uniform density sphere in solid body rotation \citep[e.g.][]{Mestel, Crampin, Dalcanton}. In contrast, our profile shapes are essentially set by radial inflow and mass conservation only (Eqn.~(\ref{eq:dimen})).

We interpret the divergence of the surface density towards the center as material that forms a bulge (including a central black hole). In order to study the bulge to disk mass ratio (B/D), defined as
\begin{equation}
\frac{\rm{B}}{\rm{D}} \equiv \frac{M_{\rm{bulge}}}{M_{\rm{bulge}}+M_{\rm{disk}}} ~~~,
\end{equation}
of our solutions, we first define the bulge radius, as the distance from the center of the disk at which the gas surface density is larger than the surface density predicted by the best fitting exponential profile by a factor of two. The calculated disk scale radii, bulge radii and the corresponding B/D for the three $u$ test-cases are summarised in Table~\ref{tab:theTable}, 
\begin{table}
\centering
\renewcommand{\arraystretch}{1.5}
\renewcommand{\tabcolsep}{0.2cm}
\begin{tabular}{|c||c|c|c|}
\hline
 Case &  $x_{\rm{disk}}$ &$x_{\rm{bulge}}$ & B/D\\
\hline
\hline
(\textit{a})  &0.28&0.075 & 0.11\\
\hline
(\textit{b}) & 0.43 &0.02 & 0.05\\
\hline
(\textit{c}) &  0.28 & 0.07& 0.11 \\
\hline
\end{tabular}
\caption{A summary the disk scale radius, bulge radius and the bulge to disk mass ratios for the three $u$-test cases discussed in the test, in steady state, with no star formation.} 
\label{tab:theTable}
\end{table}
where it can be seen that the (\textit{a}) and (\textit{c}) disks, are indeed very similar, as predicted by comparing Eqns.~(\ref{eq:yQSS}) and (\ref{eq:wow}).
Enhanced stellar feedback efficiencies could remove the accumulated ``bulge material'' near the disk centres \citep[e.g.][]{Maller, Dutton, Governato, Brook} but we do not include such feedback effects in our computations.

\subsection{Turning on star formation ($\alpha\neq 0$)}
\label{sec:YSF}
So far we have considered the gas steady state solution without any star formation or outflows. Thus the only parameter that determines the gas surface density profile is the inflow velocity. We now turn on the star formation circuit, which is also responsible for the outflowing gas from the disk back to the CGM/IGM. In this section we will focus  on cases (\textit{a}) and (\textit{b}), and will return to case (\textit{c}) only when we take the DM component into account.

In the current treatment, fixing $u$ to the two limiting cases, there is no difference between star formation and outflows. From a physical point of view, though, these are different in essence, since while outflowing gas escapes the gravitational potential of the disk, forming stars stay in the disk, and contribute to the gravitational potential, the rotational velocity and the Toomre-$Q$ parameter. We  will distinguish between those two processes in \S\ref{sec:CosEvol}.

Solving the steady-state equation ($\partial y / \partial\tau = 0$) with $\alpha\neq 0$ for cases (\textit{a}) and (\textit{b}) allows us to quantify the effect of star formation on the steady state solutions for the surface density. Although tedious, both cases are analytically solvable, and converge continuously to Eqn.~(\ref{eq:yQSS}) as $\alpha \rightarrow 0$. We present the analytic solution for case (\textit{a})  in Appendix \S\ref{sec:SSSAN0}. 

By plugging in physically motivated values for $R,v_R,\eta$ and $t_{\rm dep}$, we can get a feel of reasonable $\alpha$-values. Considering a $\sim10\,\rm kpc$ disk, inflow velocity $\left|v_R \right| \sim 20\,{\rm{km}}\,{\rm{s}}^{-1}$, depletion time of $t_{\rm{dep}}\sim 0.5 \,\rm{Gyr}$ at $z=2$ \citep{Tacconi}, and $\eta\sim2$ (see for example \citealt{Lilly}) we get $\alpha\sim3$. 

In Fig.~\ref{fig:y} (\emph{left}) we plot the steady state solutions $y_\alpha(x)$ for $\alpha=0,0.5,3$ and $10$ of case (\textit{a}). As $\alpha$ increases the gas surface density of the bulk of the disk has a shallower slope. This is because larger $\alpha$ means more efficient star formation and less efficient inflow, or $t_{\rm{crs}}>t_{\rm{dep}}$. 
In this limit the evolution at every radius $x$ is almost independent, thus resulting in a more uniform disk. For small $\alpha$, as the  depletion time becomes larger, gas flows towards the center rapidly, not staying long enough to form stars, until there is practically no star formation at all, and Eqn.~(\ref{eq:yQSS}) is relevant.
Fig.~\ref{fig:y} shows that (quasi-)exponential disks form naturally when the radial gas crossing time is short compared to the star-formation gas depletion times.

The effect of $\alpha\neq0 $ on case (\textit{b}) is similar. Moreover, even though the steady state profiles of the two $u$-cases with $\alpha=0$ are significantly different across the disk, as $\alpha$ increases both surface densities diminish, until they are barely distinguishable for $\alpha=10$.

\subsection{Convergence to steady state}
\label{sec:CCA}

Having analysed the steady state solutions of Eqn.~(\ref{eq:cont}) , we turn to solving its (dimensionless) time evolution, but still for a cosmological snapshot, i.e. assuming constant cosmological inflow rate and non-varying halo parameters. For simplicity, we focus on the $\alpha=0$ case. By carrying out this analysis, we will be able to estimate the dimensionless time $\tau$ it takes the disk to reach a steady state. As we already noted, the cosmological snapshot solution should be a reasonable approximation, if the steady state convergence timescale is short compared to the time during which the cosmological conditions change significantly.

In Fig.~\ref{fig:y} (\emph{right}) we plot the dimensionless surface density $y(x)$, for case (\textit{a}), at a few $\tau\leq1$ intervals. The gas surface density $y$ reaches a steady state by $\tau\approx1$ (the $u=-1/\sqrt{x}$ case reaches a steady state even sooner, within $\tau\approx0.75$), that is within one crossing time, converging to the analytic solution of Eqn.~(\ref{eq:yQSS}), starting off with an empty disk. Considering, for example, a $R=10\,\rm kpc$ and $v_R=20\,{\rm km}\,{\rm s}^{-1}$ disk, means that the system reaches a steady state within half a Gyr. At $z\gtrsim 2$, the DM halo (and therefore the cosmological accretion onto the disk as well) change significantly within half a Gyr. At later times, however, the cosmological accretion rate changes only slightly, meaning that a quasi steady state is maintained.

In addition, we also find that the convergence to the steady state surface density is an ``outside-in" process. As a consequence, the fact that the radial surface density profile is well approximated by an exponential profile is not true throughout the evolution of the disk. This approximation becomes better as the evolution approaches its quasi steady state.

\section{Cosmological evolution}
\label{sec:CosEvol}

So far we have considered halos of constant mass, and thus constant baryonic accretion rate, to which we referred to as a cosmological snapshot. As we have shown, within this picture the disk evolution equation can be written in a purely dimensionless form and solved analytically. We now generalise our model by accounting for the evolution of those cosmological conditions as well as other galactic parameters and local star formation, which is a necessary step for the model to be self consistent: 

\begin{figure*}
\centering
\includegraphics[width=\columnwidth]{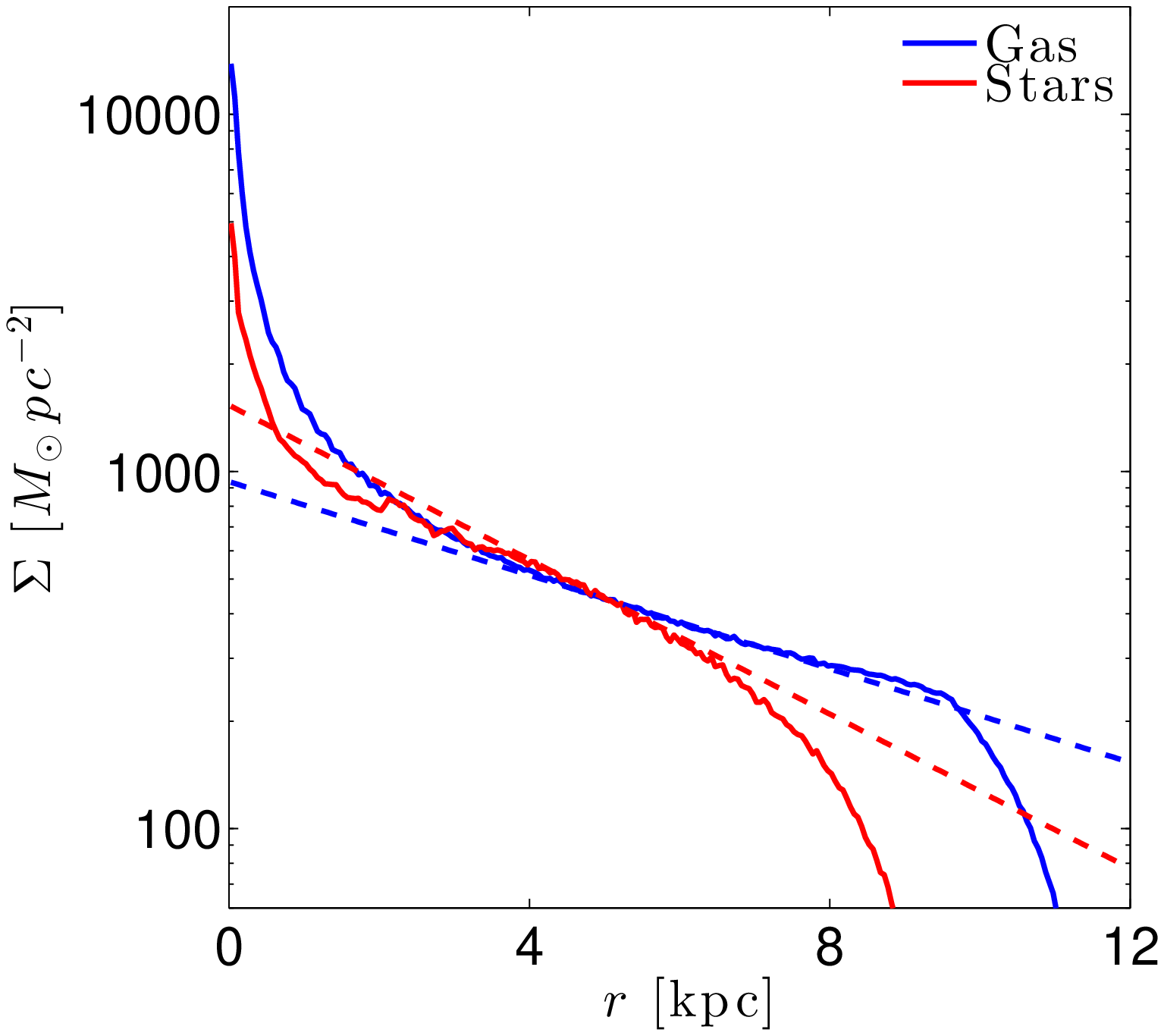}
\includegraphics[width=\columnwidth]{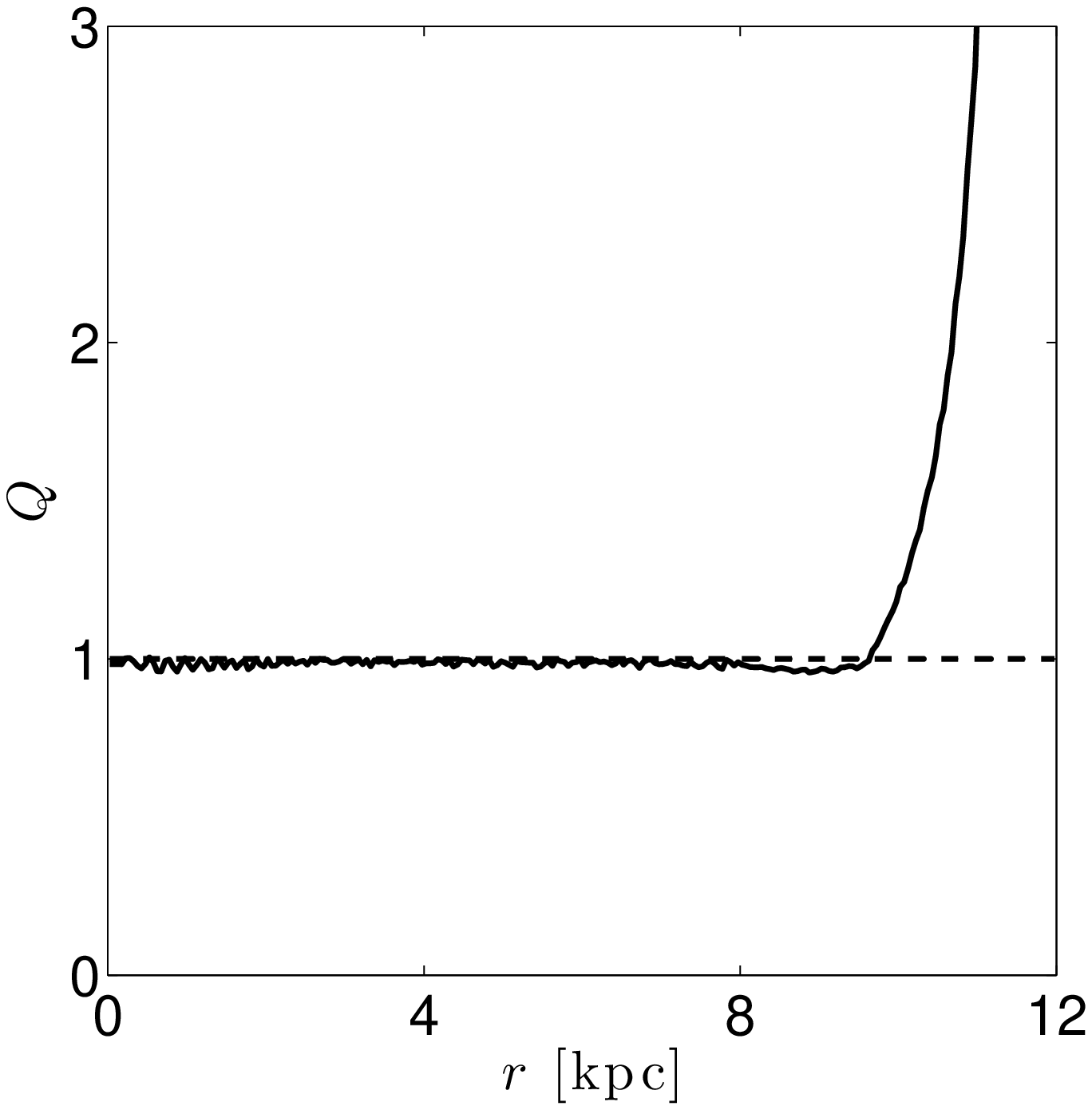}
\includegraphics[width=\columnwidth]{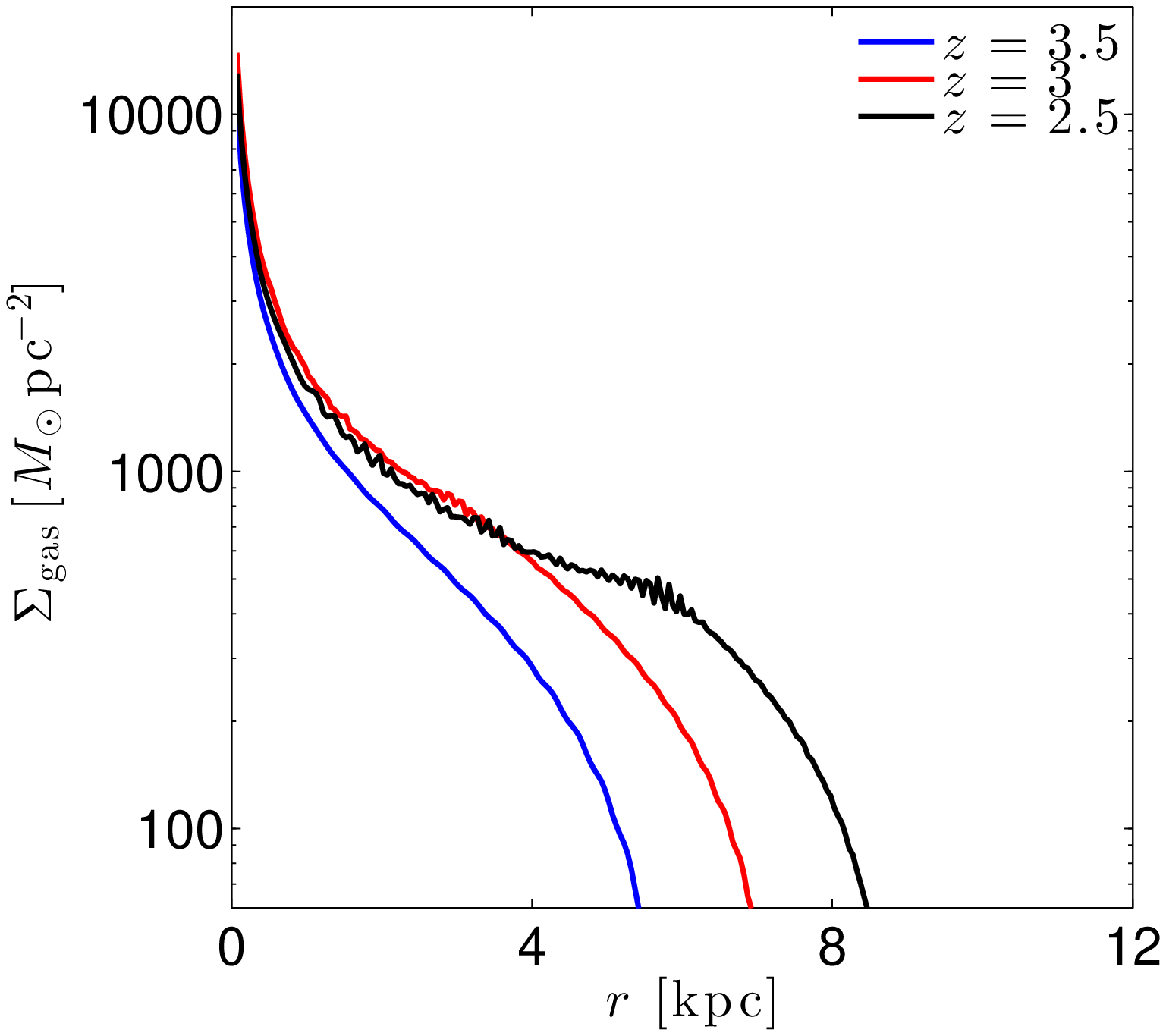}
\includegraphics[width=\columnwidth]{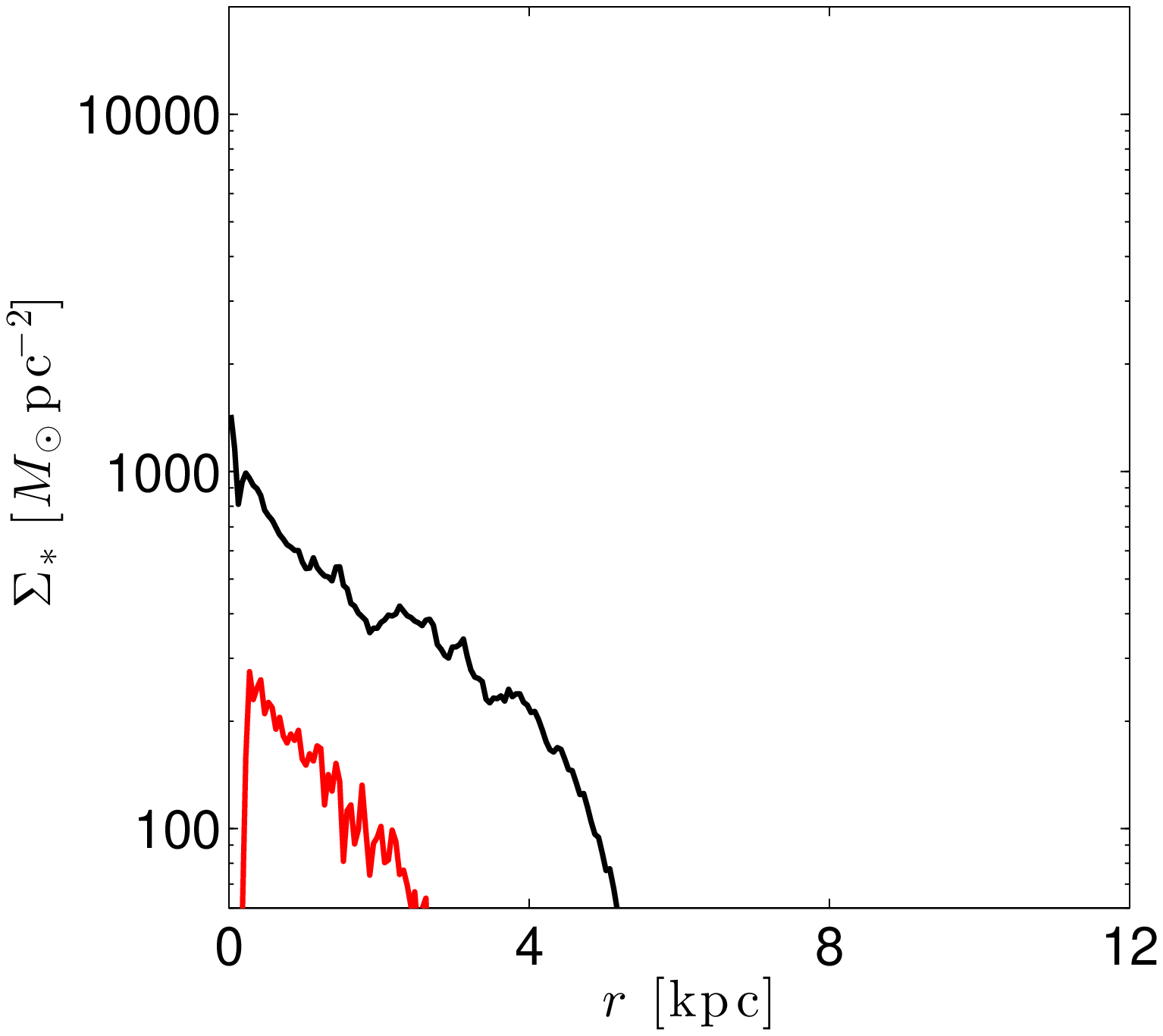}
\caption{ \emph{top left}: a numerical solution to the mass-continuity equation (\ref{eq:dimen}) for the fiducial model --- for the gas (blue) and for the stars (red) --- as a function of the in plane distance from the center of the disk. (\emph{top right}): the calculated toomre-$Q$ parameter, at $z=2$, as a function of the distance from the center of the disk, for the fiducial model, together with the star formation threshold. According to our model, star formation is taking place everywhere in the disk that satisfies $Q(r)\leq1$. \emph{bottom}: The evolution of the gaseous \emph{left} and stellar (\emph{right}) profiles, from $z=3.5$ (cosmic time $\sim1.8\,{\rm Gyr}$) until $z=2.5$ ($t\sim 2.6\,{\rm Gyr}$).}
\label{fig:fid}
\end{figure*}

\begin{itemize}
\item \textbf{DM halo} --- 
The DM halo is a crucial ingredient in  disk formation and evolution for a couple of reasons. First, the halo growth rate, that depends on $M_h$ sets the baryonic accretion rate onto the disk. In our model we follow \citet{Dekel-2,Dave-2}, taking the baryonic accretion rate as $\dot{M}_{\rm in}=\dot{M}_h f_B/(1-f_B) $, where the halo growth rate is 
\begin{equation}\label{eq:Dekel}
\dot M _h = 390 \left( \frac{M_h}{10^{12}M_\odot}\right) ^{1.15}  \left( \frac{1+z}{3}\right) ^{2.25} M_\odot\, {\rm yr}^{-1} ~~~.
\end{equation}
and can be integrated to give the exact halo mass at redshift $z$, assuming that its mass at some $z_0$ is known, as we show in Appendix \S\ref{sec:MhOfZ}.

Second, the DM halo also affects the rotation curve of the baryonic disk. As a result not only the mere total halo mass is important, but also its spatial distribution $M_h(r)$. Furthermore, due to the dependence of the rotation curve on the halo mass distribution, the inflow velocity and the disk size will also be affected. Therefore, to account for the mass distribution correctly, we must specify the halo concentration $\mathcal{C}$ as a function of $M_h$ and $z$. We assume continuously evolving median concentration halos following \citet{Prada}. We neglect adiabatic contraction of the dark matter halos

\item \textbf{Sound speed} --- 
Since in our model the gas inflow velocity is $\propto c_s^2$ and $Q\propto c_s/\Sigma$, the gas sound speed is an essential parameter. Higher sound speed means gas that flows inwards more efficiently, resulting in low $\Sigma$-values at large $r$, with the addition of a larger numerator in $Q(r)$. Therefore, we expect that the larger the gas sound speed is, the harder it gets for the disk to form stars, and the smaller the stellar disk becomes.

We follow the redshift evolution of the sound speed, as in \citet{Wisnioski}, i.e. 
\begin{equation}\label{eq:cs}
c_s=18(1+z)\,{\rm km}\,{\rm s}^{-1}~~~. 
\end{equation}
As noted in \citet{Wisnioski}, this relation is valid over a wide range of redshifts. However, due to measurement uncertainties, $c_s$ shows a substantial scatter around this value (for example $\pm10\,{\rm km}\,{\rm s}^{-1}$ at $z=0$, and $\pm30\,{\rm km}\,{\rm s}^{-1}$ at $z=2$). Therefore, we will use Eqn.~(\ref{eq:cs}) as our fiducial sound speed in \S\ref{sec:fid}, and  study how our model predictions change when considering different $c_s$ in \S\ref{sec:BFM}, keeping the $1+z$ dependence on redshift.

\item \textbf{Disk size} --- To constrain the disk size, we assume that the specific angular momentum of the accreted matter, $(J/M)_{\rm accr}$, equals that of the parent DM halo, $(J/M)_{\rm halo}$, where the accreted gas is assumed to be in centrifugal equilibrium with the disk, i.e. it starts rotating around the center of the disk at a rotational velocity $v_{\rm rot}$ immediately as it enters the disk. Not making this assumption, necessarily means that other assumptions, about the time it takes the newly accreted gas to reach the disks rotational velocity and the amount of energy that is dissipated in this process, have to be made. It would also cause a situation in which at each radius $r$, the gas rotation is inhomogeneous. Since we assume that the cosmological accretion at each time is uniform across the disk, then the constraint on the conservation of the angular momentum reads
\begin{equation}\label{eq:AMConstraint}
\left(\frac{J}{M} \right)_{\rm halo} =\left(\frac{J}{M} \right)_{\rm accr} \equiv \frac{\int_0^R r^2v_{\rm rot} \mathrm{d}r }{\int_0^R r \mathrm{d}r } ~~~.
\end{equation}
Therefore for a given $\lambda$, since we know the halo mass at each redshift, we can solve Eqn.~(\ref{eq:AMConstraint}) for the disk size $R_d=R_d(z)$. 
With Eqn.~(\ref{eq:AMConstraint}) we are essentially adopting the results of both observations (e.g. \citealt{Burkert2}) and cosmological hydrodynamic simulations \citep[e.g.][]{GenelNew, Danovich} that show that specific angular momentum is conserved from halo to disk. This despite the different angular momentum histories of the gas and DM. Thus, we only specify the spin parameter $\lambda$, but not the cumulative mass distribution of the specific angular momentum, $M_h(<j)$, found in dark matter only simulations as cosmological initial conditions \citep[e.g.][]{Bullock,Dutton2}.

\item \textbf{Depletion time} --- 
At the base of our radially resolved equilibrium model lies the equilibrium relation (\ref{eq:EQ}), that relates the SFR to the gas content of the disk. The proportionality, $t_{\rm dep}$, quantifies the efficiency of star formation --- the shorter $t_{\rm dep}$ is, the more efficient star formation becomes, which can proceed even if the gas in the disk flows rapidly inwards due to a high inflow velocity. For the analysis below, we take 
\begin{equation}\label{eq:tdep}
t_{\rm dep} = (1+z)^{-0.34} \, {\rm Gyr} ~~~, 
\end{equation}
following \citet{Genzel}. 
\item \textbf{Star formation} --- 
It has already been noted by \citet{Agertz, Ceverino,Cacciato, Genel} that $Q$ in the disk will self-regulate to maintain $Q\sim1$. The intuitive explanation for this self-regulation in \citet{Cacciato} is generally valid in the spatially resolved case as well. Nevertheless, resolving the disk radially, means that $\kappa,\,\Sigma$ and $Q$ are all calculated as a function of the distance from the center of the disk. Wherever the ratio between the epicyclic frequency  and $\Sigma$ is small enough, $Q$ will reach its threshold value, and self regulation will proceed. At the outer regions of the disk, however, the low gas surface density will dominate $Q(r)$, which will consequently be $>1$, and therefore star formation will not be initiated 
as expected for the outer lower surface density portions of galaxy disks \citep[e.g.][]{Schaye, vanDerKruit, Clark}. 
This does not imply that regions in the disk where no star formation is taking place at some redshift will remain quiescent. However,  due to the cosmological evolution of the parent halo, and as a result of the baryonic accretion rate and the size of the disk, star formation can begin at a later time. 

In order to account for star formation self-consistently in the analysis below, we calculate $Q(r)$, and turn the star formation-circuit on only at those regions of the disk in which $Q<1$.
\end{itemize}

\section{Results}
\label{sec:results}
In this section we consider cosmological evolution of the halo and the accretion rate, employ a self consistent star formation criterion, and solve numerically the evolution equations for the gas and the stars in the disk. First, in \S\ref{sec:fid}, we define and analyze thoroughly our fiducial model. Next, in \S\ref{sec:BFM} we examine how our results depend on model parameters, $\lambda,c_s$ and $\eta$ for a fixed (at $z=2$) halo mass. Finally, in \S\ref{sec:BFM2} we study how the model results change when the halo mass is modified.

\begin{figure*}
\centering
\includegraphics[width=\columnwidth]{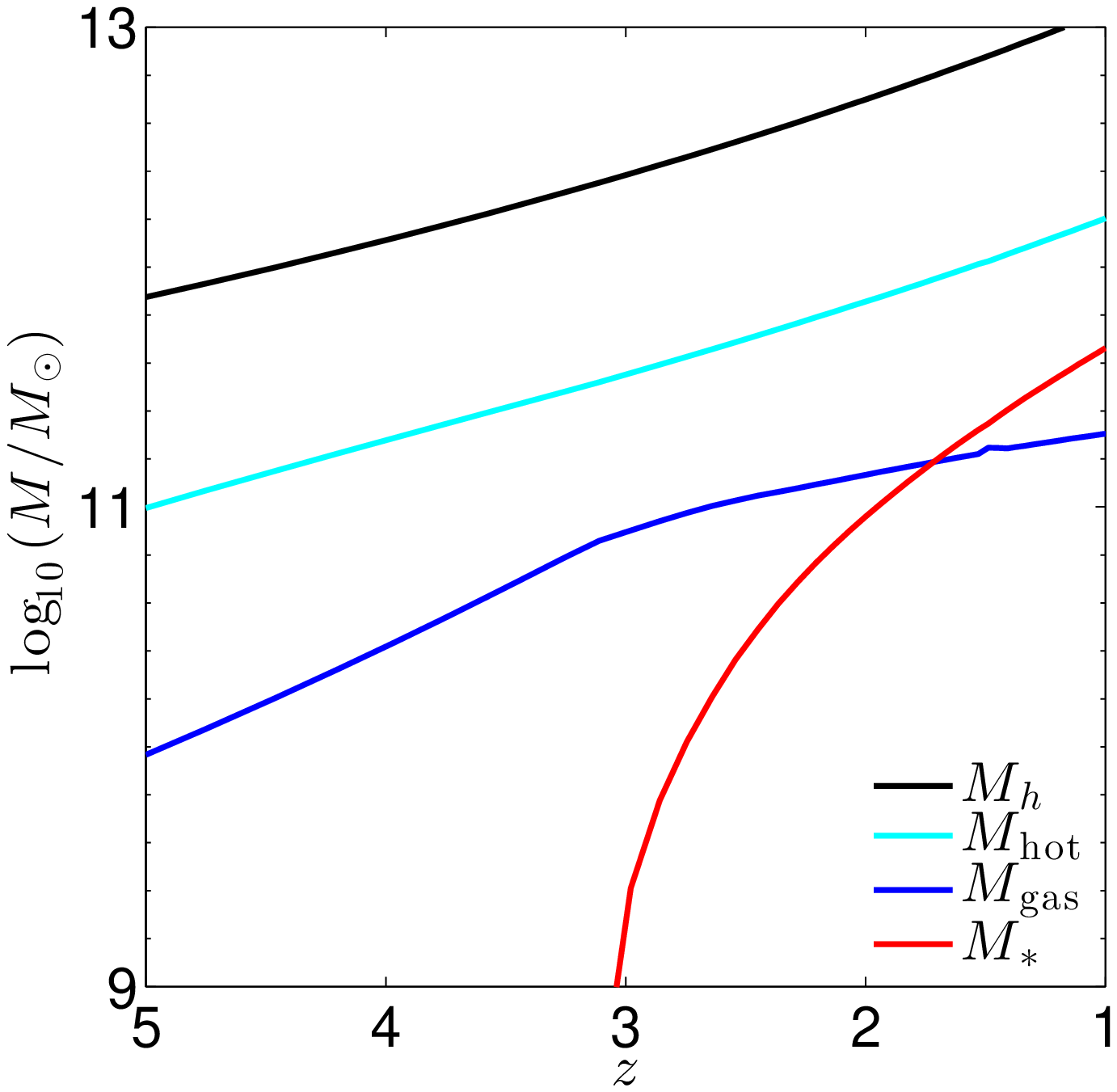}
\includegraphics[width=\columnwidth]{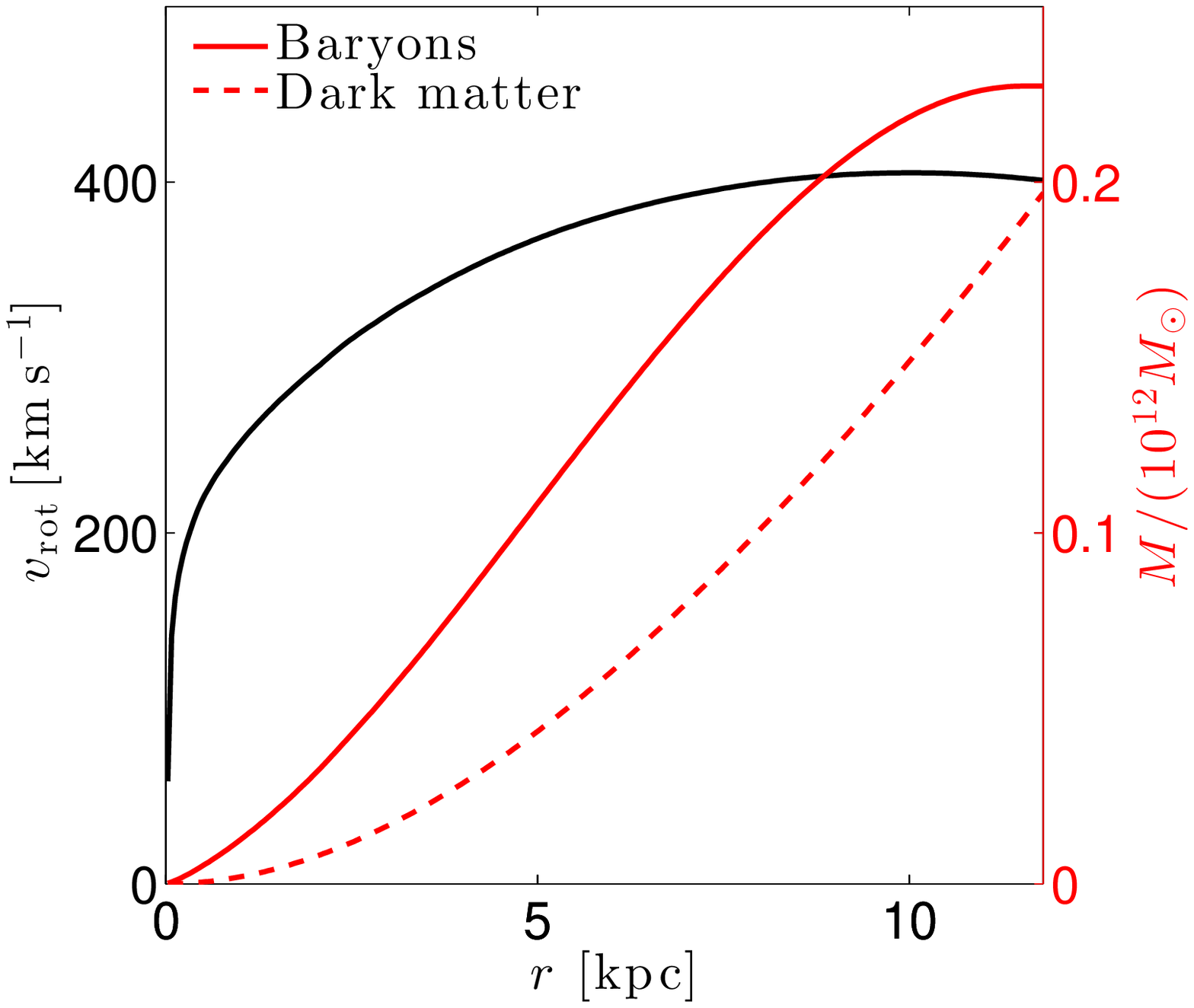}
\caption{(\emph{left}): Redshift evolution of the total mass in the halo, hot gas, cold gas and stars  in the numerical solution for the evolution of our fiducial model. (\emph{right}): Radially resolved $v_{\rm rot}$ (\emph{black}) and baryonic and DM mass contributions (\emph{red}) for the fiducial model at $z=2$.}
\label{fig:msOfZ}
\end{figure*}

\subsection{Fiducial model}
\label{sec:fid}
As our fiducial model we assume a $\Lambda$CDM cosmology, with $\Omega_m=0.3$, $h=0.7$, and baryon fraction $f_b=0.16$ \citep{Planck}. We also assume that the DM halo profile follows a Navarro-Frenk-White profile \citep{NFW} with median concentration parameter that varies with redshift, following \citet{Prada}. The fiducial DM halo we consider is of mass $M_h=5\times 10^{12}M_\odot$ at $z=2$ and spin parameter $\lambda=0.035$. In addition, we assume $\eta=2$, no instantaneous recycling, and $c_s=54\,{\rm{km}}\,{\rm s}^{-1}$ at $z=2$  within the disk. Our massive fiducial halo represents the tip of the galaxy main-sequence at $z\approx2$.

Solving for the cosmologically time-dependent fiducial halo, in which the star formation term is turned on only when and where $Q<1$, we plot in Fig.~\ref{fig:fid} (\emph{top left}) the resulting surface density of the two disk components at $z=2$, and the corresponding $Q(r)$ in Fig.~\ref{fig:fid} (\emph{top right}). As can be seen in the figure, for the fiducial model both gas and the stellar profiles are well approximated by exponential profiles (dashed lines), for a wide range of radii. 
For the stars and gas distributions, exponential profiles with scale lengths equal to $6.6$ and $4\,\rm{kpc}$ respectively, represent the computed distributions within a factor of two, from 0.5 to 9 $\rm kpc$ for the stars, and from 1 to 11 $\rm kpc$ for the gas.

Remarkably, though somewhat fortuitously, the stellar length scale precisely equals the length scale given by $\lambda r_{\rm vir} /\sqrt{2}$ for assumed exponential disk in an isothermal halo \citep{MMW}.

According to our bulge-criterion, the fiducial model bulge extends to $\sim0.85\,\rm kpc$, and the bulge to disk mass ratio is ${\rm B/D} = 0.05$. In addition, while the gaseous disk extends roughly to $12\,\rm kpc$, the stellar disk is somewhat more compact, stretching out to $9\,\rm kpc$. The gas half mass surface density, defined as $\Sigma_{\rm gas}(r_{1/2,\rm gas})$ is $400~M_\odot\,{\rm pc}^{-2}$, where $r_{1/2,\rm gas}$ is the the distance form the center of the disk such that 
\begin{equation}
M_{\rm gas}(r_{1/2,\rm gas}) = M_{\rm gas}(R_d)/2~~~.
\end{equation}
The stellar half mass surface density, defined in a similar way is $500~M_\odot\,{\rm pc}^{-2}$. At $z=2$ we find $r_{1/2,*} = 4.5\,\rm kpc$, consistent with observed sizes for this stellar mass and redshift \citep{Natascha,Dutton}. The star formation rate (at $z=2$) for the fiducial model is $\sim 120 M_\odot\,{\rm yr}^{-1}$, and the gas fraction is $60\%$. 

Comparing our predicted SFR to the SFR-main sequence, we find that the predicted SFR for our fiducial halo, $120 M_\odot\,{\rm yr}^{-1}$, is in very good agreement with the SFR according to \citet{Whitaker}, $125 M_\odot\,{\rm yr}^{-1}$, for the same redshift $z=2$, and  stellar mass $\sim9\times 10^{10}M_\odot$.

Using our model we can also follow the actual evolution of the disk, as a function of redshift.  In Fig.~\ref{fig:fid} (\emph{bottom row}) we plot the surface density of the gaseous (\emph{left}) and stellar disks (\emph{right}) for $z=3.5, 3$ and $2.5$. It can be seen that even though that the characteristic gas surface density reaches its ``final" ($z=2$) value already at $z\sim3$, the disk growth continues even at smaller redshifts. Studying the stellar disk surface density evolution, we find that star formation begins in the inner regions of the disk at $z\sim3$, and propagates outwards with time, until at $z=2$ the stellar disk is almost double in size with respect to $z=2.5$, or $\sim0.7\,{\rm Gyr}$ later.

In Fig.~\ref{fig:msOfZ} (\emph{left}) 
we plot the redshift evolution of the total mass for each of the disk components (gas and stars), together with the DM halo and hot gas for our fiducial model, where for the hot gas we sum all the gas that flows through the disk and out from the center, plus the accumulated outflow that is proportional to the total stellar mass.  Most of the baryons remain as hot halo gas throughout the evolution. At $z\sim3$, the onset of star formation in the fiducial model case, the disk gas mass growth slows down, as is to be expected from the equilibrium relations. The effect on the hot gas mass, however, is insignificant since most of the hot gas is ejected from the disk center.

\begin{figure}
\centering
\includegraphics[width=\columnwidth]{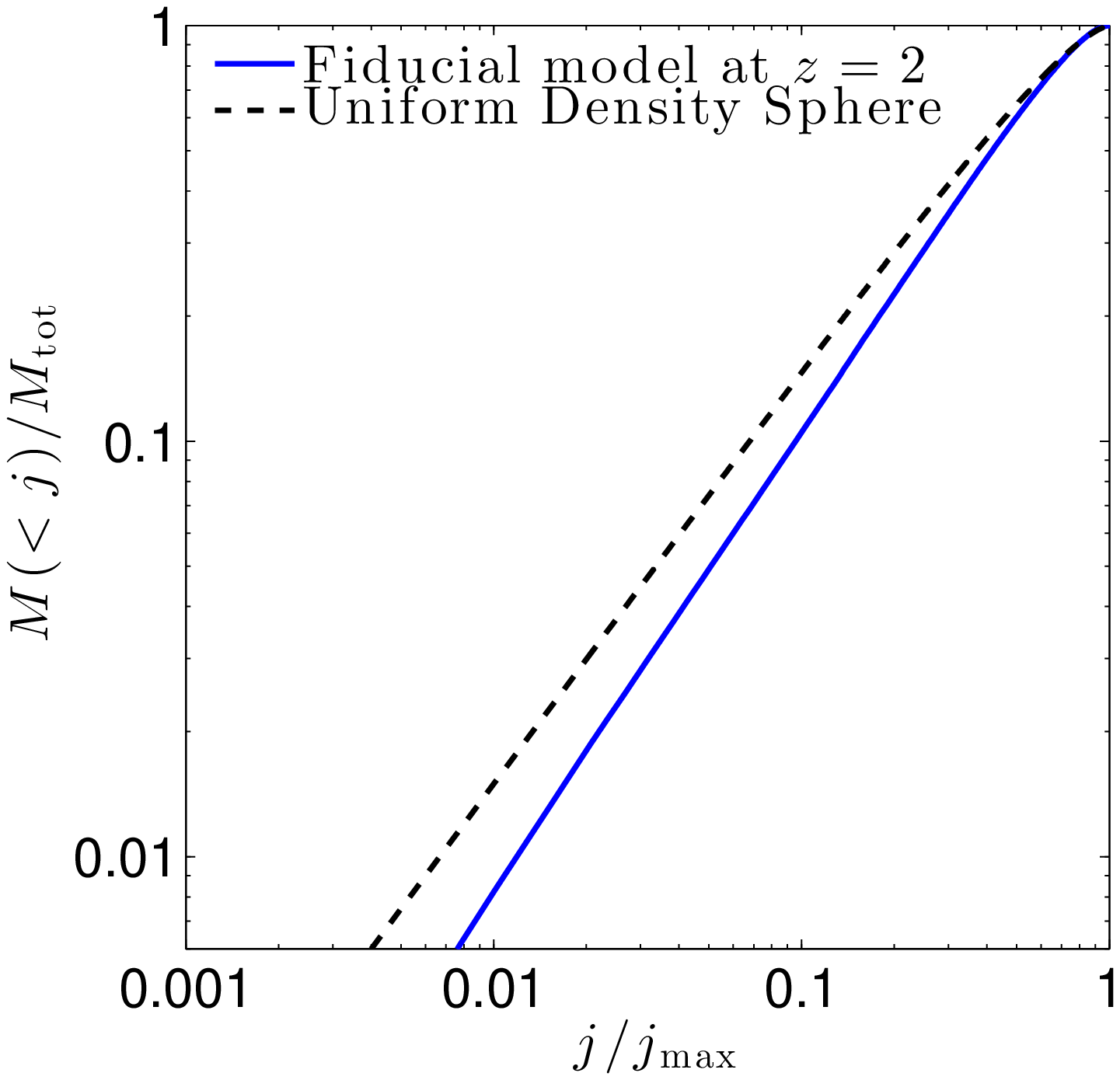}
\caption{ $M(<j)/M_{\rm tot}$ versus $j/j_{\max}$ for our fiducial model (\emph{solid blue}) and $M(<j)/M_{\rm tot} = 1 - \left(1-j/j_{\max}\right)^{3/2}$ (\emph{dashed black}) for a sphere in solid body rotation.}
\label{fig:mLessThanJ}
\end{figure}

In Fig.~\ref{fig:msOfZ} (\emph{right}) we plot the rotation curve $v_{\rm rot}(r)$ (black curve) together with the baryonic and dark matter contributions to the total mass (solid and dashed red curves). The disk is baryon dominated throughout.  It is also of interest to plot the resulting cumulative mass versus specific angular momentum, $M(<j)$, for our fiducial model.  In Fig.~\ref{fig:mLessThanJ}
we plot $M(<j)/M_{\rm tot}$ versus $j/j_{\max}$ for our fiducial model (solid curve) together with the relation $M(<j)/M_{\rm tot} = 1 - \left(1-j/j_{\max}\right)^{3/2}$ for a sphere in solid body rotation. The two curves differ only slightly. This is consistent with our resulting (quasi)-exponential disks, and the known similarity between the angular momentum distributions of centrifugally supported exponential disks and rotating spheres \citep[e.g.][]{Mestel,expNew,Dalcanton,Bullock}.

\subsection{Parameter study at fixed $M_h$ and $z$}
\label{sec:BFM}

\begin{figure*}
\centering
\includegraphics[width=0.97\columnwidth]{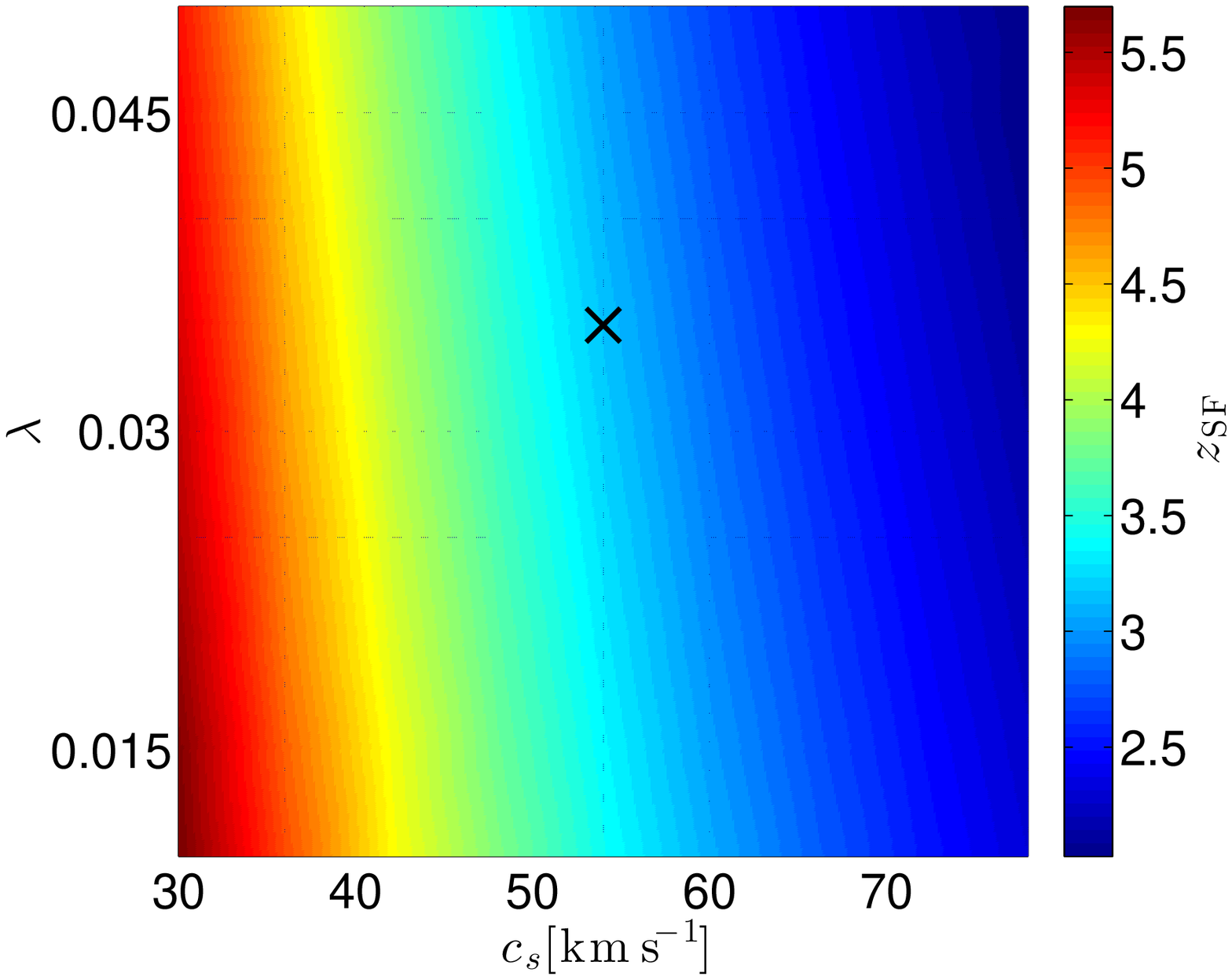}
\includegraphics[width=\columnwidth]{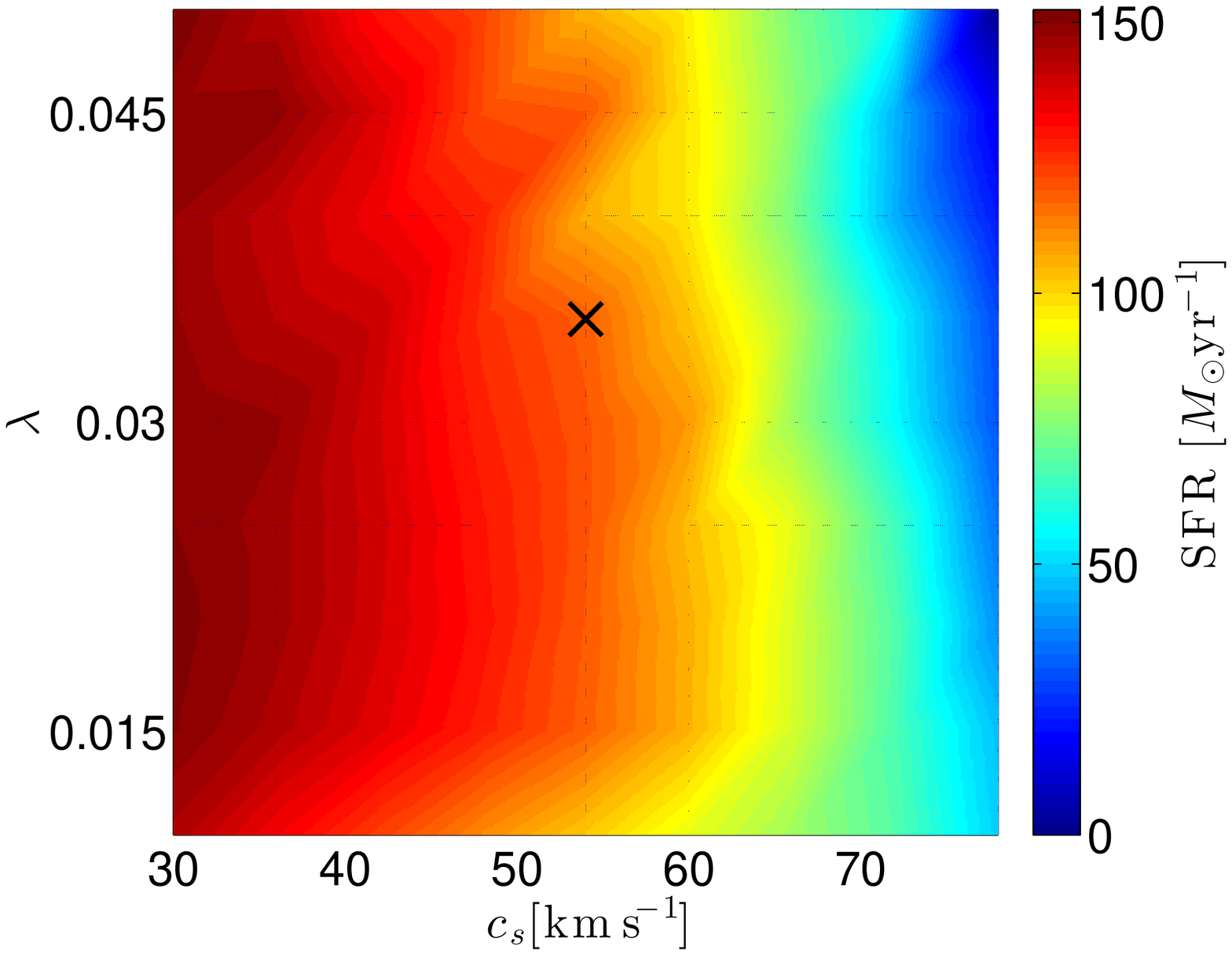}
\includegraphics[width=\columnwidth]{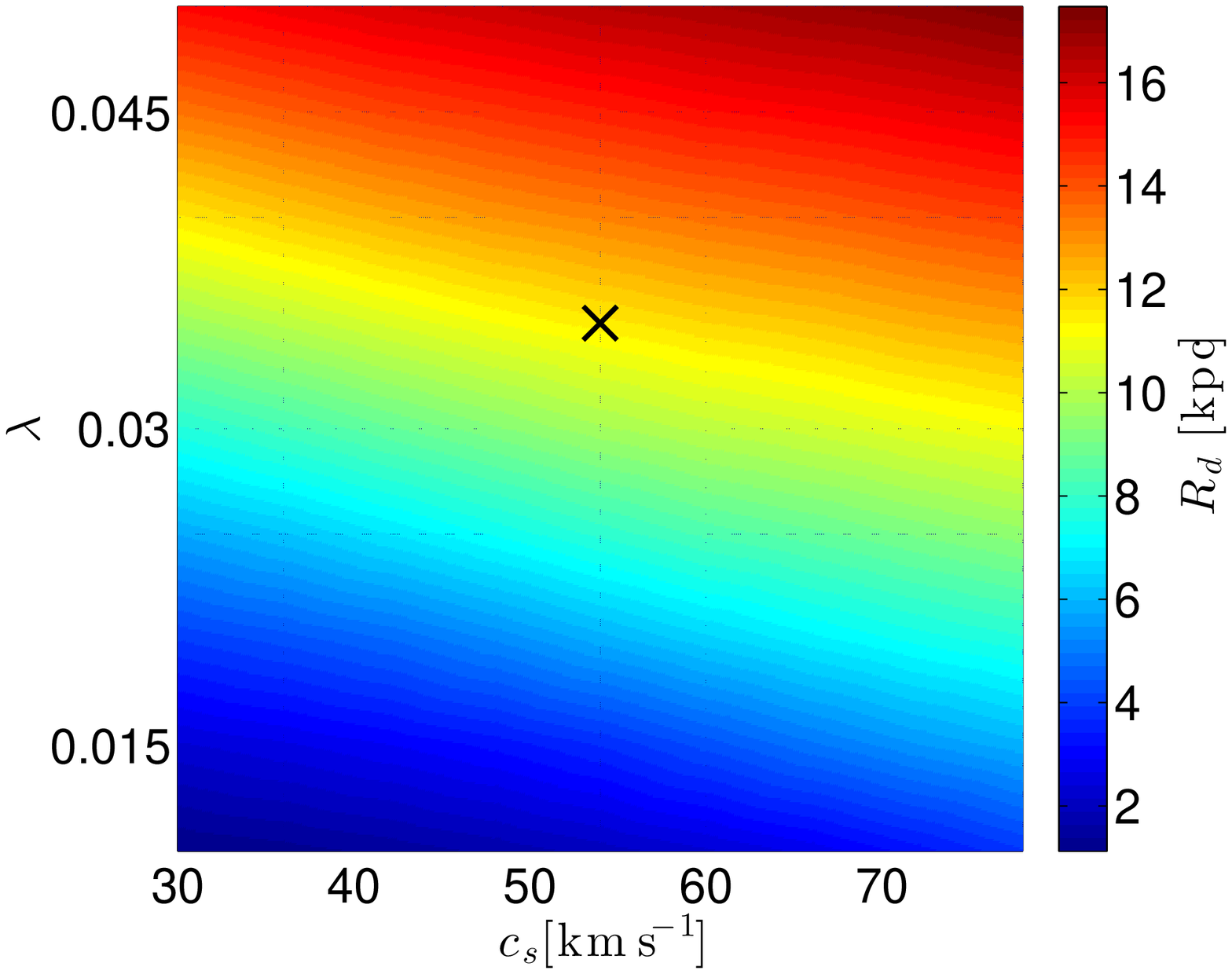}
\includegraphics[width=\columnwidth]{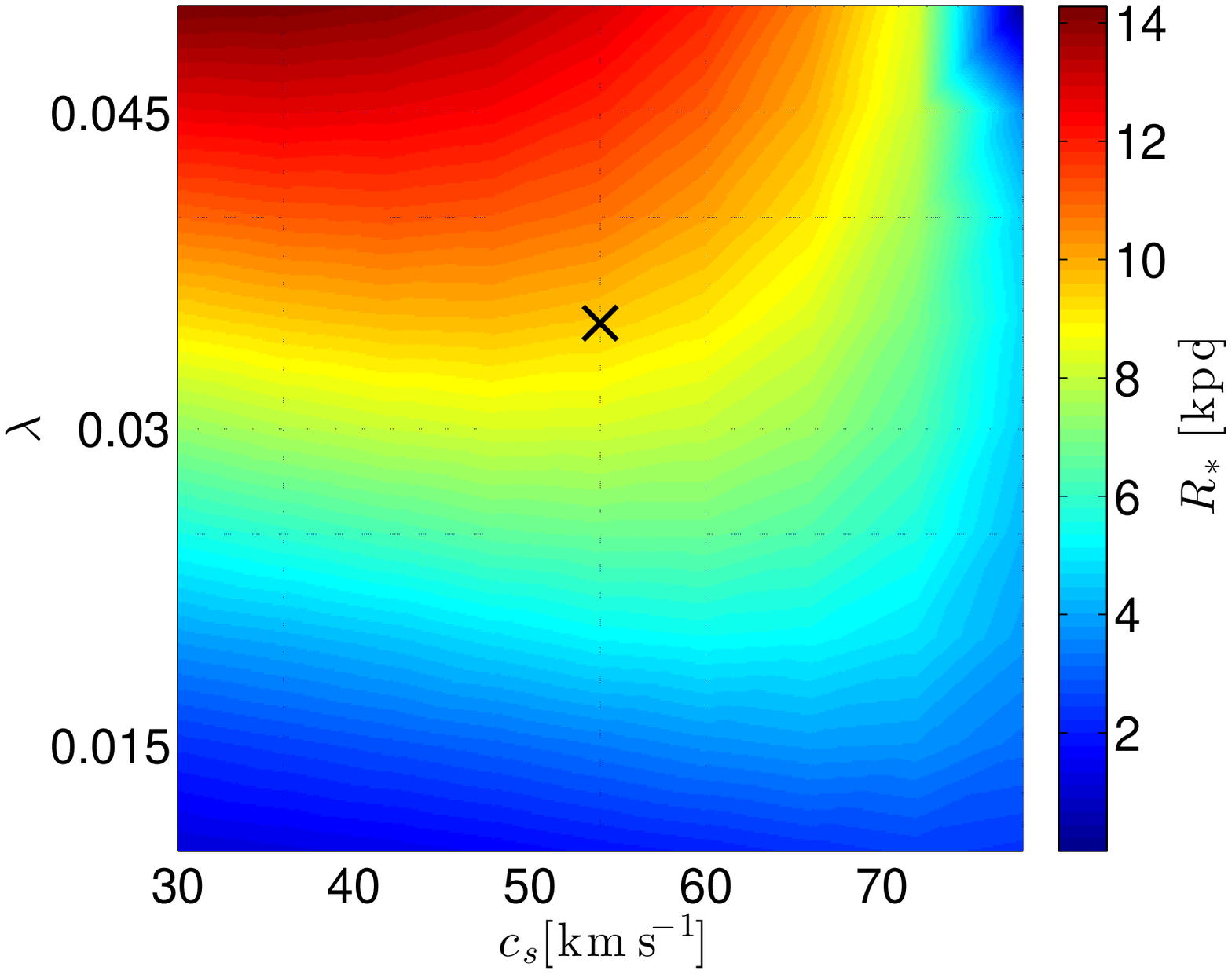}
\caption{The $z=2$ dependence on $c_s$ and $\lambda$ of (\emph{top left} to \emph{bottom right}): $z_{\rm SF}$, SFR, $R_d$ and $R_*$. }
\label{fig:cs_l_first}
\end{figure*}

\begin{figure}
\centering
\includegraphics[width=\columnwidth]{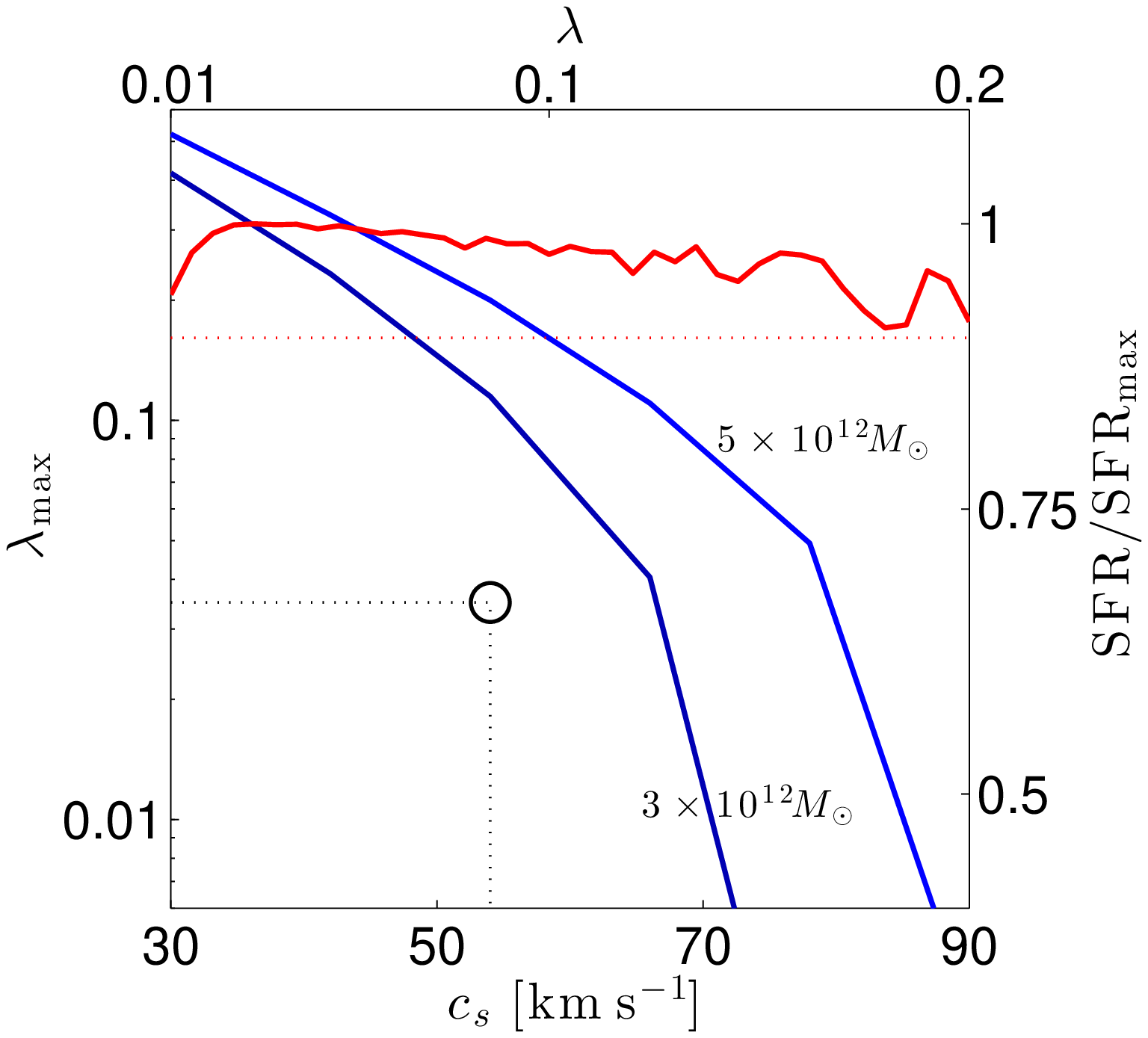}
\caption{\emph{Bottom-left axes:} $\lambda_{\max}$ that allows star formation at $z=2$ for a $5\times10^{12}M_\odot$ (\emph{blue}) and a $3\times10^{12}M_\odot$ (\emph{dark-blue}) halo as a function of the gas sound speed. The fiducial model ($\lambda=0.035,c_s=18\,{\rm km}\,{\rm s}^{-1}$) is designated by a small black circle. Only the regions below the curves are star forming. No star formation occurs above the blue curves.
\emph{Top-right axes:} $\lambda$ dependence of the total SFR for the $c_s=18(1+z)\,{\rm km}\,{\rm s}^{-1}$ case. It can be seen that the SFR dependence on $\lambda$ is weak, changing by less than $10\%$ within the $\lambda<\lambda_{\max}$ range.}
\label{fig:lambdaMax}
\end{figure}

We wish to examine how the disk properties are modified when the halo or disk parameters are altered. We note that a key advantage of our approach --- a simple toy-model of disk evolution --- is that it enables scanning through a large parameter-space at a low computational cost. We focus on $5\times10^{12}M_\odot$ halos at $z=2$, as for our fiducial model.

We examine how the redshift at which star formation begins, $z_{\rm SF}$, depends on $c_s$ and $\lambda$. Since the initial conditions for our disk evolution are of a median halo and an empty disk that is fed by cosmological accretion, star formation cannot take place until enough gas  has been accreted and $\Sigma$ is large enough. The growing disk size, on the other hand, makes $\Sigma$ smaller, thus delaying the onset of star formation.  The impact of having a large $c_s$ is threefold. First, the large inflow rate empties the gas at each radius more efficiently, resulting in reduced $\Sigma$. Second, it delays star formation due to the $Q$-barrier it induces. Third, as we show below, although counter-intuitive, larger $c_s$ results in larger disks, and hence smaller $\Sigma$, thus delaying star formation. Even though larger $\lambda$ can also delay star formation, through the disk size, for  $\lambda$ from $0.01$ to $0.05$, which we consider, its effect is weaker, as seen in Fig.~\ref{fig:cs_l_first} (\emph{top left}). 

Interestingly, after star formation has begun, $\lambda$ hardly has an effect on SFR, despite its effect on the disk size. Qualitatively, increasing $\lambda$ means increasing the angular momentum of the halo, and hence also of the accreted baryons. Therefore, as $\lambda$ increases, the disk size $R$  increases too. However, since the baryonic accretion onto the disk 
\begin{equation}
\dot{M}_{\rm in} \propto M_h^{1.15}\left( 1+z \right)^{2.25} ~~~,
\end{equation}
is independent of $\lambda$, then to first order the total baryonic mass in the disk will also be independent of $\lambda$. Consequently, even though large  $\lambda$ means smaller characteristic $\Sigma$, as long as it is high enough to maintain $Q\sim1$, star formation, which is $\propto M_{\rm gas}$ should proceed at more or less the same rate, as seen in Fig.~\ref{fig:cs_l_first} (\emph{top right}).

Nevertheless, there should exist a $\lambda_{\max}$, that once reached, renders $\Sigma$ too low and $Q$ too high, which will not allow any more star formation. In Fig~\ref{fig:lambdaMax} (\emph{bottom-left axes}) we plot $\lambda_{\max}$ that allows star formation at $z=2$, as a function of the gas sound speed, for our fiducial halo ($5\times10^{12}M_\odot$ at $z=2$), and for a slightly less massive halo ($3\times10^{12}M_\odot$). It can be seen that $\lambda\sim0.035$, as preferred by observations and simulations is well within the star forming portion of the parameter space, and becomes borderline only as the gas sound speed increases significantly. In Fig~\ref{fig:lambdaMax} (\emph{top-right axes}) we plot the total SFR for the fiducial sound speed, as a function of $\lambda$. It can be seen that for $\lambda$ altered by more than an order of magnitude, the total SFR changes by less than $10$ per cent, following our qualitative argument.

Studying the disk size, one might expect that since $v_r\sim c_s^2$, then for higher sound speeds, the gaseous disk will be more compact, as the radial migration of the gas becomes more efficient. However, the disk size is primarily set by the baryonic accretion onto the disk, whose extent, in our model, is set by conservation of angular momentum. Let us consider a halo of mass $M_h$, with some specific angular momentum, $(J/M)_{\rm halo}$ that is, of course, independent of the sound speed. A larger $c_s$ at the early stages of the disk evolution, will result in a mass profile that is more concentrated near the center of the disk, and thus lower baryonic specific angular momentum. Therefore, to satisfy the angular momentum constraint, Eqn.~(\ref{eq:AMConstraint}), the disk must grow for higher sound speeds.
\begin{figure*}
\centering
\includegraphics[width=\columnwidth]{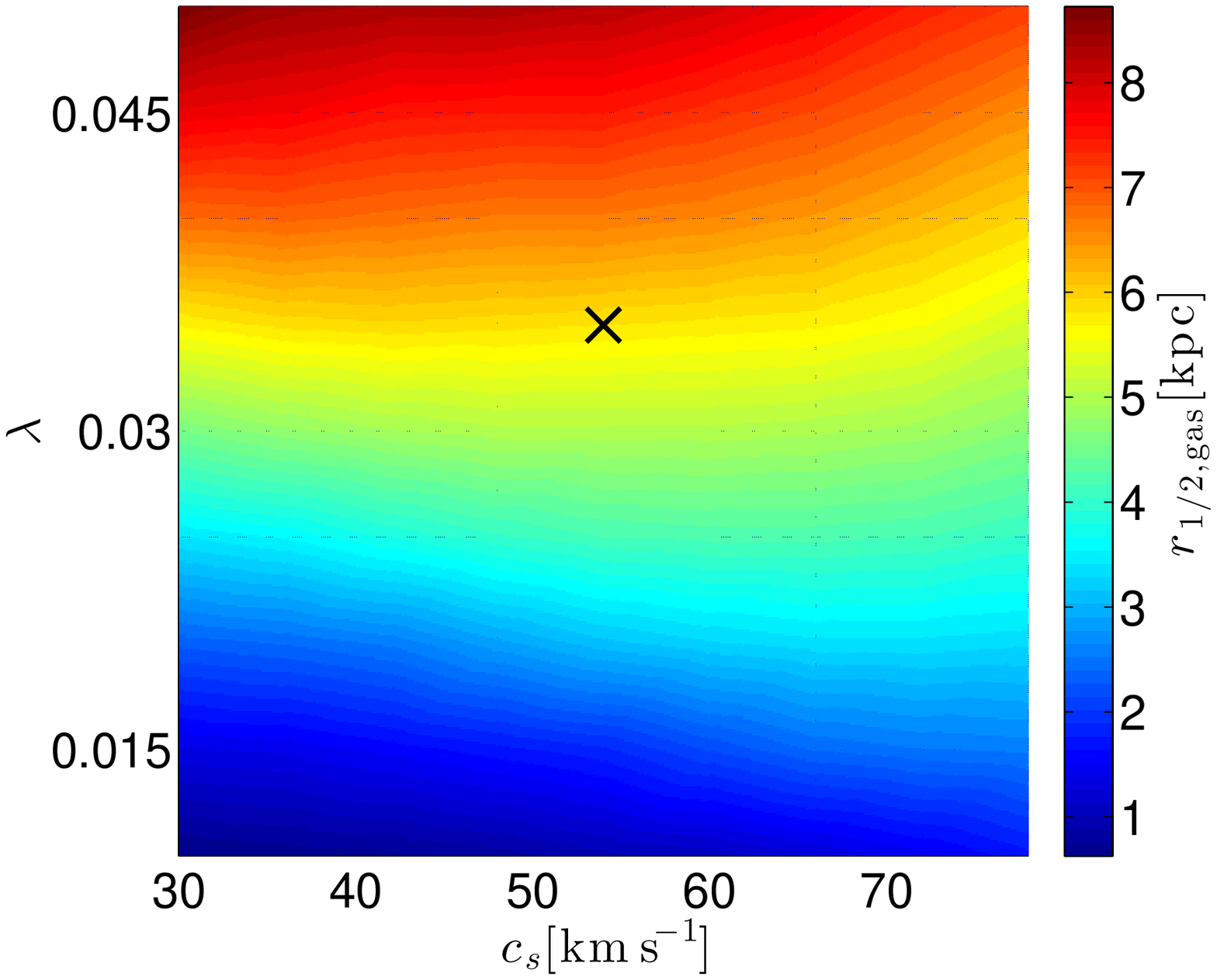}
\includegraphics[width=\columnwidth]{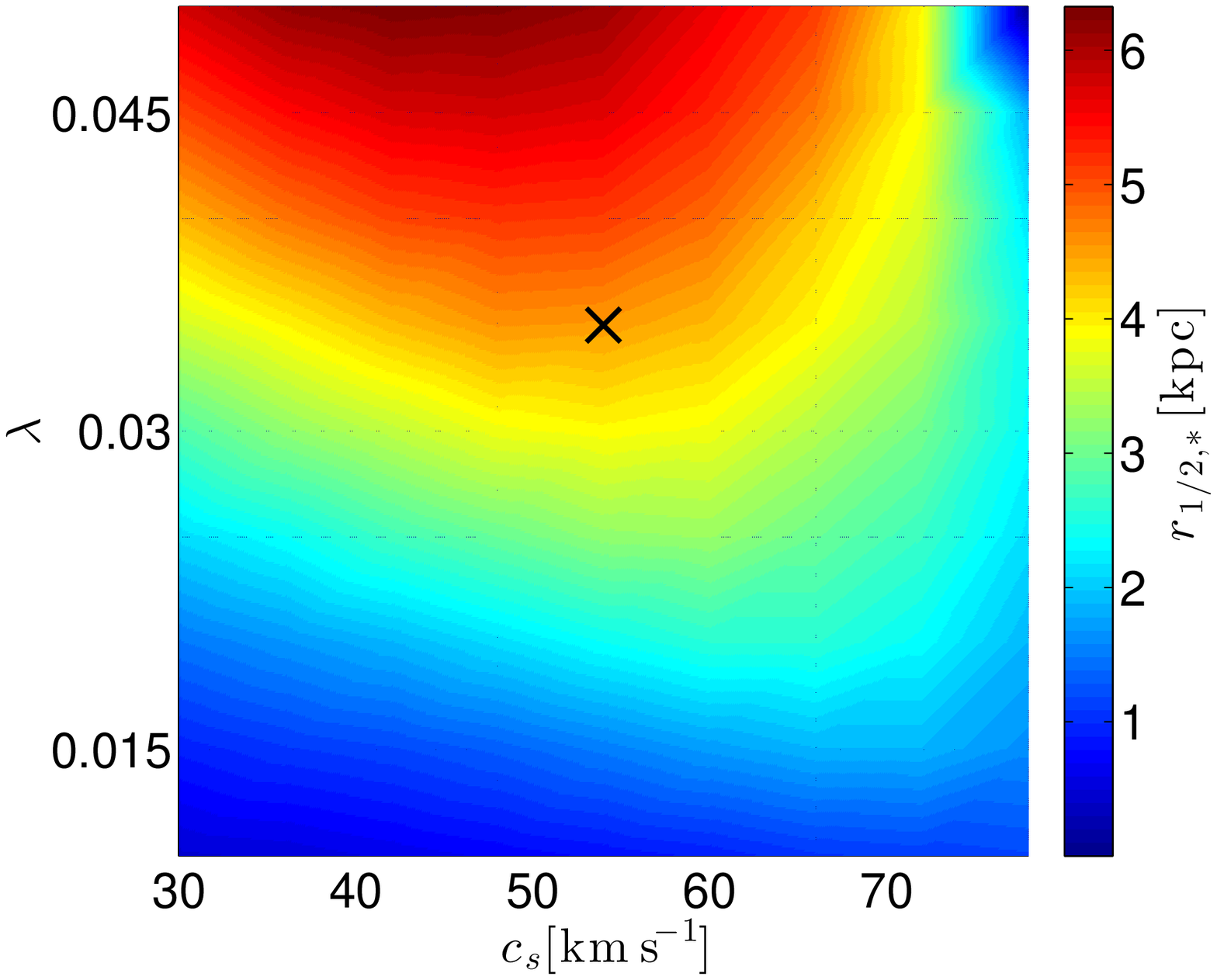}
\includegraphics[width=\columnwidth]{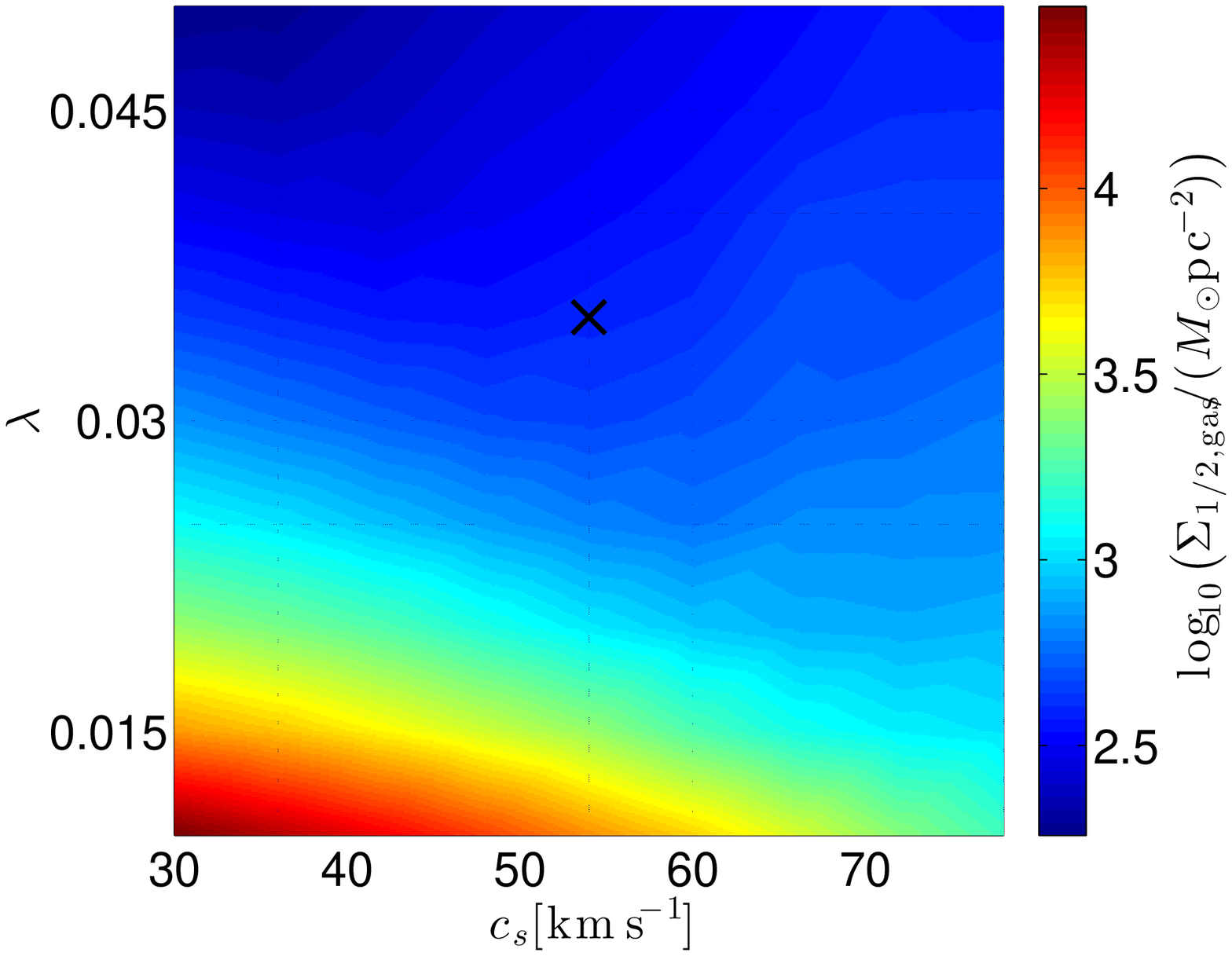}
\includegraphics[width=\columnwidth]{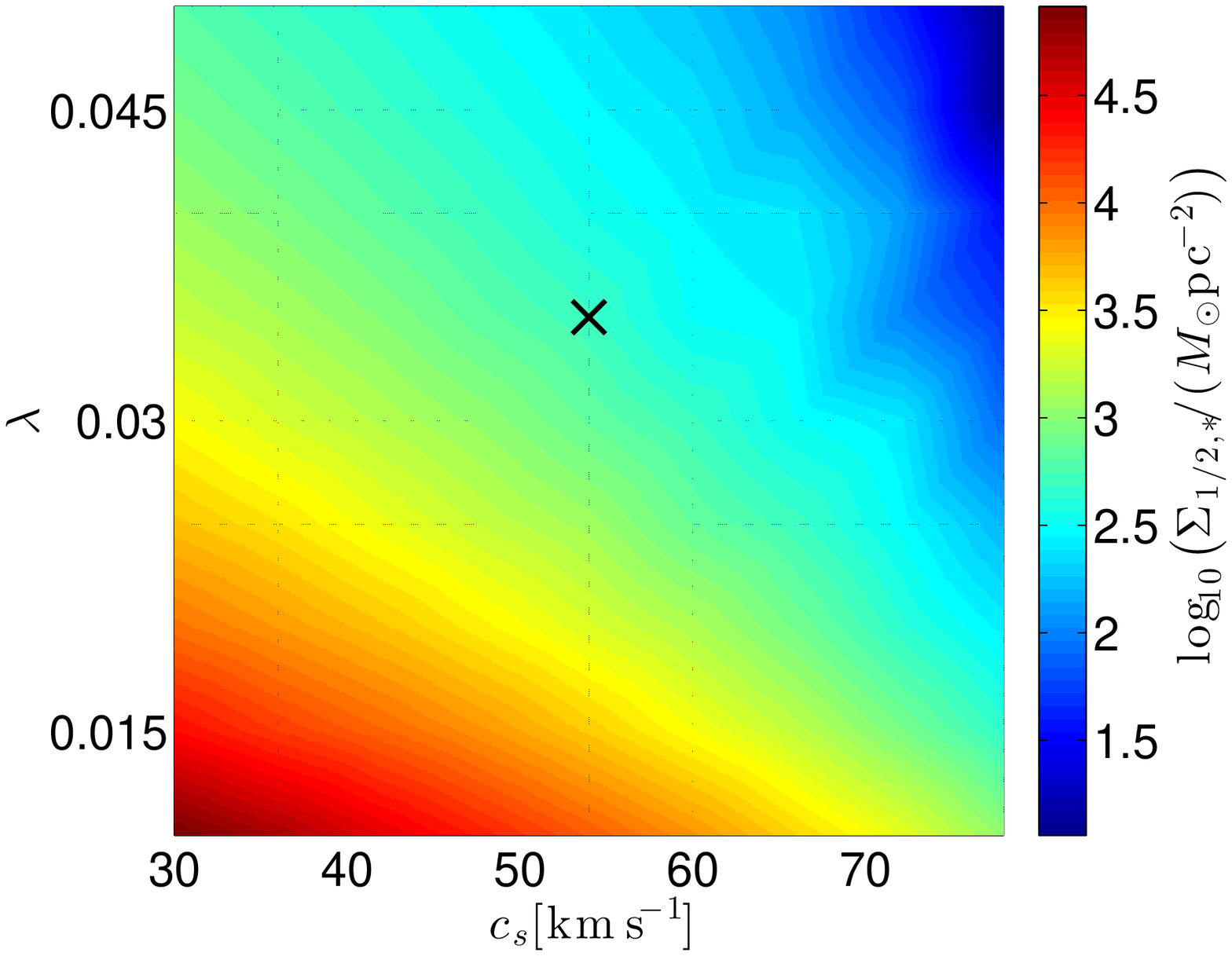}
\caption{The $z=2$ dependence on $c_s$ and $\lambda$ of (\emph{top left} to \emph{bottom right}): $r_{1/2,\rm gas}, r_{1/2,*}, \Sigma_{1/2,\rm gas}$ and $\Sigma_{1/2,*}$. }
\label{fig:cs_l_More}
\end{figure*}
This is seen in Fig.~\ref{fig:cs_l_first} (\emph{bottom left}), where we plot $R_d$ (the truncation radius of the baryonic disk) as a function of the gas sound speed and $\lambda$, for our fiducial halo at $z=2$. 

For the  stellar disk size dependence on $c_s$, which is plotted in Fig.~\ref{fig:cs_l_first} (\emph{bottom right}), we note that there are two competing effects. As $c_s$ increases, Eqn.~(\ref{eq:Q}) implies that in the outer regions of the disk, where $\Sigma$ is lower, star formation is suppressed. But, since the gaseous disk grows as $c_s$ grows, then the absolute size of the stellar disk is potentially larger, at least as long as overall $Q$ is small enough. 

In Fig.~\ref{fig:cs_l_More}  we plot the corresponding $r_{1/2}$ (\emph{top row}) and $\Sigma_{1/2}$ (\emph{bottom row}), the half mass radius and surface density of the gas (\emph{left column})  and stars (\emph{right column}) . It can be seen that regions in parameter space that result in small $R_d$ and $R_*$ have large $\Sigma_{1/2}$. This is because small disks get a larger mass contribution from the central divergence of the surface density. The opposite is not exactly satisfied -- the largest disks do not have the lowest half mass surface densities. For the gas component, larger disks contain more baryons, and hence their characteristic surface density is higher, while for the stellar component not all of the parameter space results in disks that form stars, resulting in lower characteristic surface densities.

\begin{figure*}
\centering
\includegraphics[width=\columnwidth]{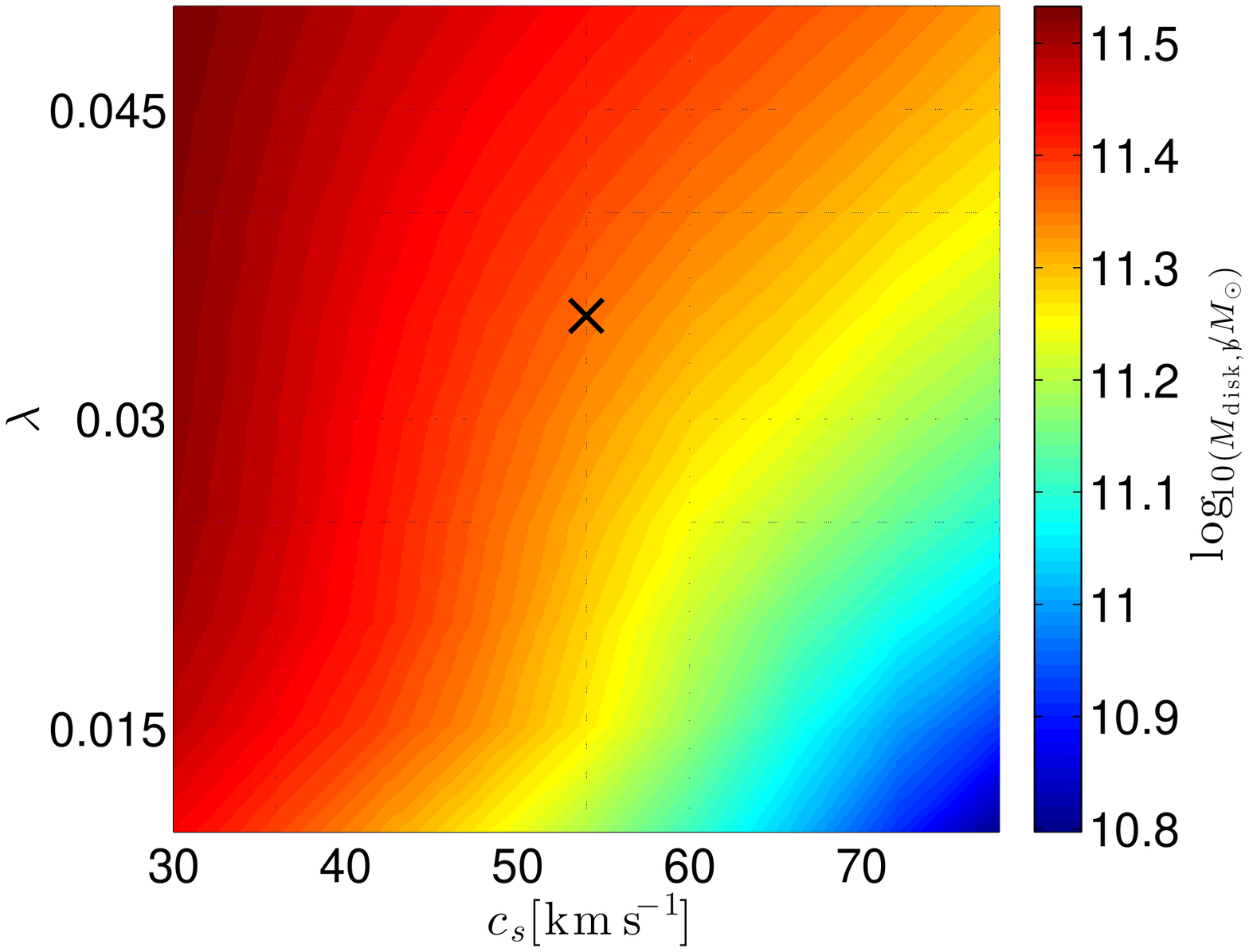}
\includegraphics[width=\columnwidth]{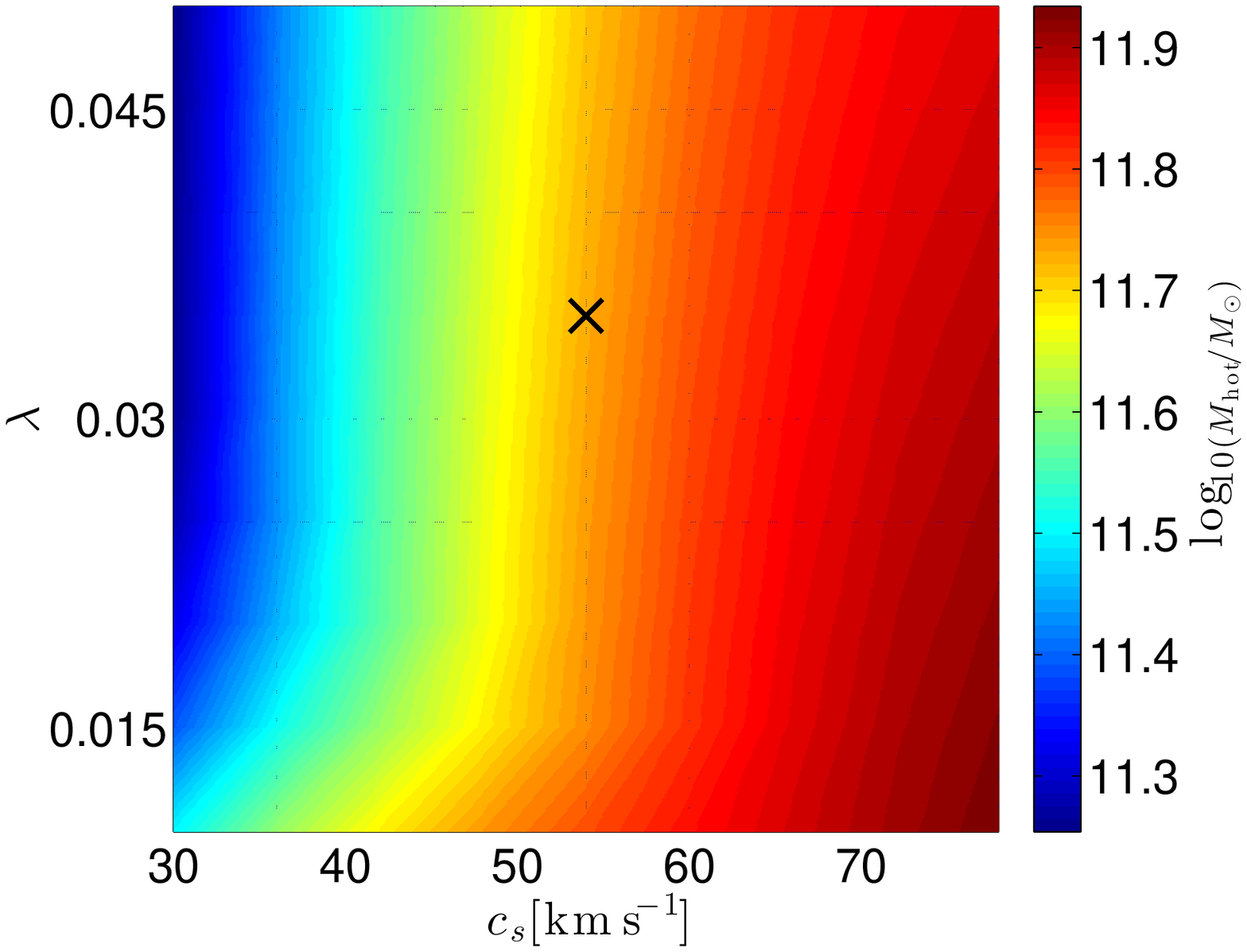}
\includegraphics[width=\columnwidth]{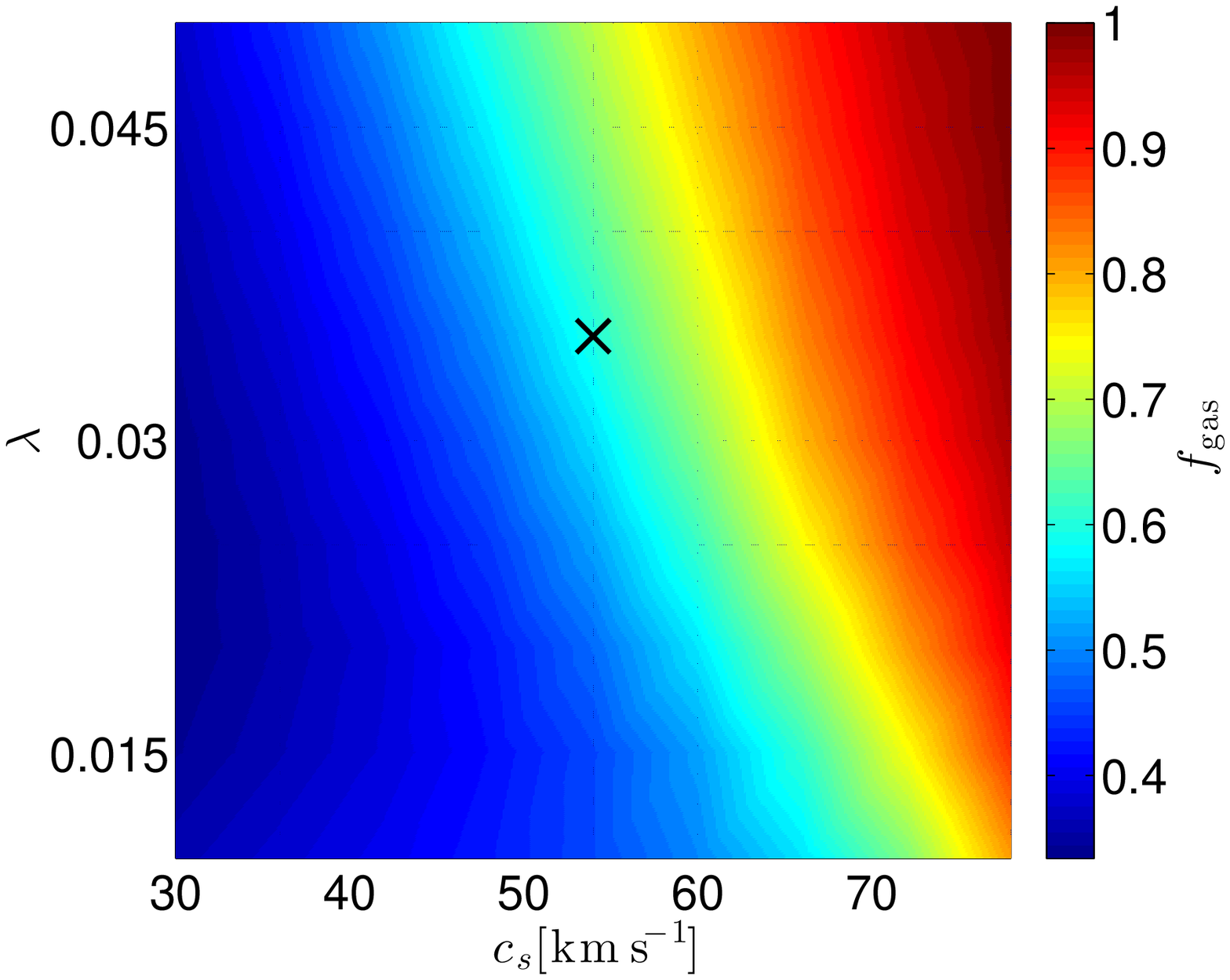}
\includegraphics[width=\columnwidth]{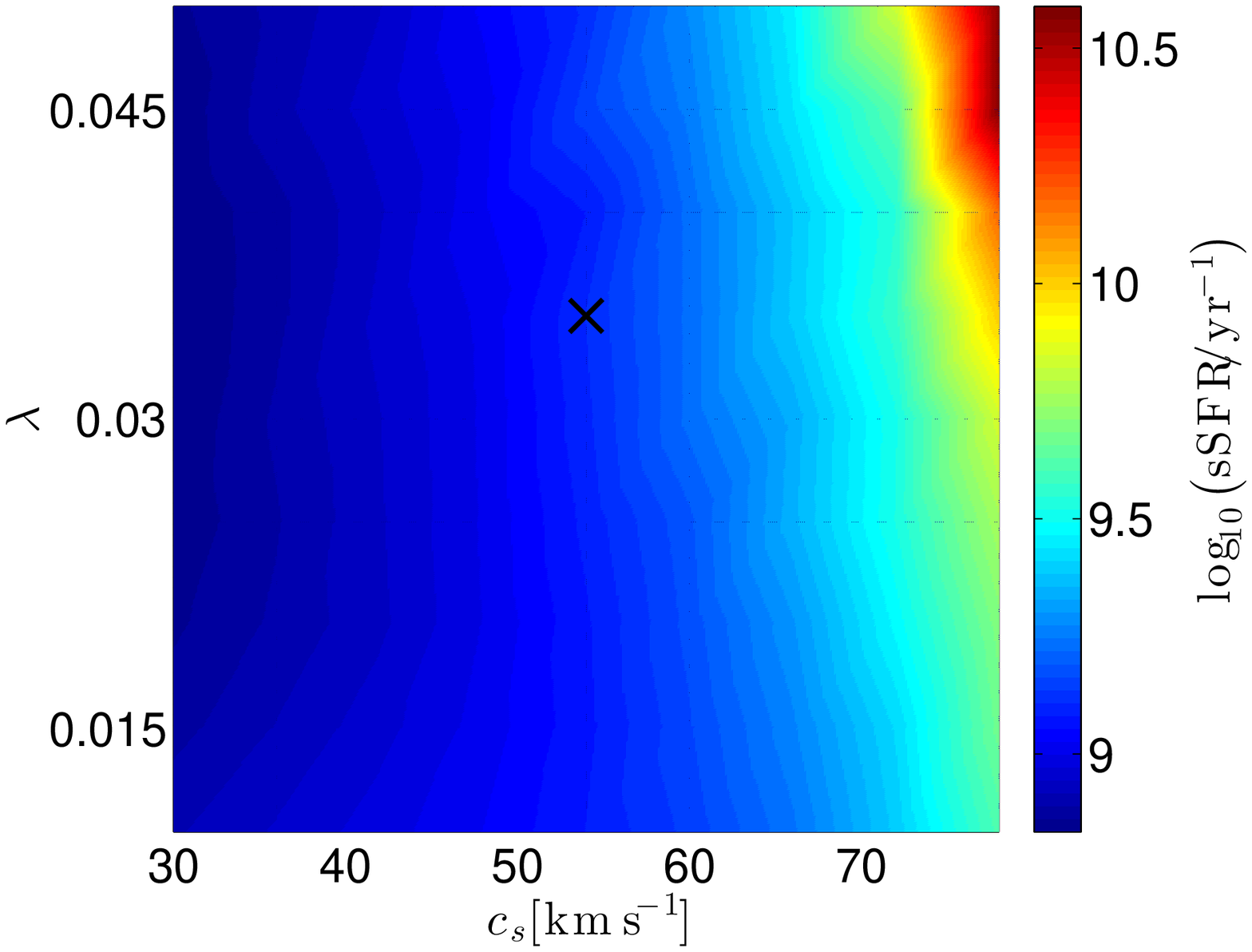}
\caption{The $z=2$ dependence on $c_s$ and $\lambda$ of (\emph{top left} to \emph{bottom right}): $M_{\rm disk}, M_{\rm hot}, f_{\rm gas}$ and sSFR. }
\label{fig:cs_l_MsAndMore}
\end{figure*}

Owing to these effects, the total disk mass, as well as the relative gaseous and stellar contributions to it depend on $c_s$ and $\lambda$ in a non-trivial way.  As seen in Fig.~\ref{fig:cs_l_MsAndMore} (\emph{top left}), the total disk mass increases as the sound speed decreases, as one would expect, primarily due to the fact that higher $c_s$ results in a rapidly emptying disk, where the excess gas that inevitably reaches the center of the disk and ekected back to the hot halo. The hot halo is also fed by the outflowing baryons, whose mass $\propto$ SFR. The total hot gas mass due to these contributions, is plotted as a function of $c_s, \lambda$ for our fiducial halo at $z=2$ in Fig.~\ref{fig:cs_l_MsAndMore} (\emph{top right}). It can be seen that the hot gas mass depends very weakly on $\lambda$, similarly to the SFR. This implies that outflows constitute a significant part of the total hot gas mass. The relative contributions of gas and stars to the total disk mass can be read off Fig.~\ref{fig:cs_l_MsAndMore} (\emph{bottom left} and \emph{bottom right}), 
where we plot the gas fraction $f_{\rm gas}$, and $ {\rm sSFR}={\rm SFR}/M_* $, respectively. It can be seen that except for the fact that the gas fraction is lower where SFR is higher, it exhibits a stronger dependence on $\lambda$ than the SFR, since as argued above higher $\lambda$ generates larger disks,  that results in a smaller portion of it that forms stars. 

Apart from $\lambda$ and $c_s$, another free parameter in our model is the mass loading factor $\eta$. Considering variation in $\eta$ from $1$ to $3$ results in insignificant modifications to the results reported above, and we thus do not pursue this $\eta$ analysis further. We note, however, that analysis of the extreme case $\eta=0$, i.e. no outflows at all, shows a stronger dependence of SFR and $M_{\rm hot}$ on $\lambda$, verifying our argument about the importance of outflows to the hot gas mass.

\subsubsection{SFR versus $M_*$}

\begin{figure}
\centering
\includegraphics[width=\columnwidth]{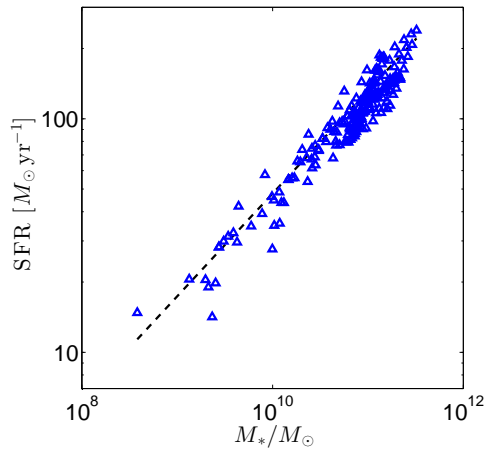}
\caption{SFR-$M_*$ relation (or ``main sequence'', as predicted by our model for various model parameters ($\lambda, c_s, \eta$) at $z=2$, and the corresponding observed relation (dashed) according to \citet{Whitaker}. }
\label{fig:SFRScatter}
\end{figure}

As we have already mentioned, the tight scaling relation between the SFR in disk galaxies and their stellar mass is remarkable. This scaling relation has been repeatedly observed, in many independent observations and different surveys. Therefore, it is interesting to compare our model predictions to actual observations of SFR, and how well those predictions fall within the galaxy main sequence. In particular, we compare our model predictions to the best-fitting SFR-$M_*$ relation, as found in \cite{Whitaker}, for star forming galaxies between $z=0$ and $2.5$
\begin{equation}\label{eq:whit}
\log_{10}\rm{SFR} = \alpha(z)(\log_{10}M_*-10.5)+\beta(z) ~~~,
\end{equation}
where
\begin{eqnarray}
\alpha(z)&=&0.7-0.13z \nonumber \\
\beta(z)&=&0.38+1.14z-0.19z^2 ~~~.
\end{eqnarray}
Since the initial conditions of our simulated disks are of empty disk, and since, as we have shown, in our model star formation is initiated at $z\sim3$, it is clear that the correspondence between our model predictions and Eqn.~(\ref{eq:whit}) cannot hold for any high redshift, and thus we are interested in comparing these SFR's at a lower redshift, so that the disk has already reached an equilibrium. In Fig.~\ref{fig:SFRScatter} 
we plot the Whitaker relation, Eqn.~(\ref{eq:whit}), as a function of $M_*$ for $z=2$ (\emph{black dashed}) compared to our model predictions, all at $z=2$ and for a halo of mass $5\times10^{12}M_\odot$, and various model parameters, $c_s,\lambda$ and $\eta$. It can be seen from this plot that our model predicts very well the SFR-$M_*$ relation, even as $c_s, \lambda$ and $\eta$ are varied.

\subsection{Varying $M_h$ at $z=2$}
\label{sec:BFM2}
\begin{figure*}
\centering
\includegraphics[width=\columnwidth]{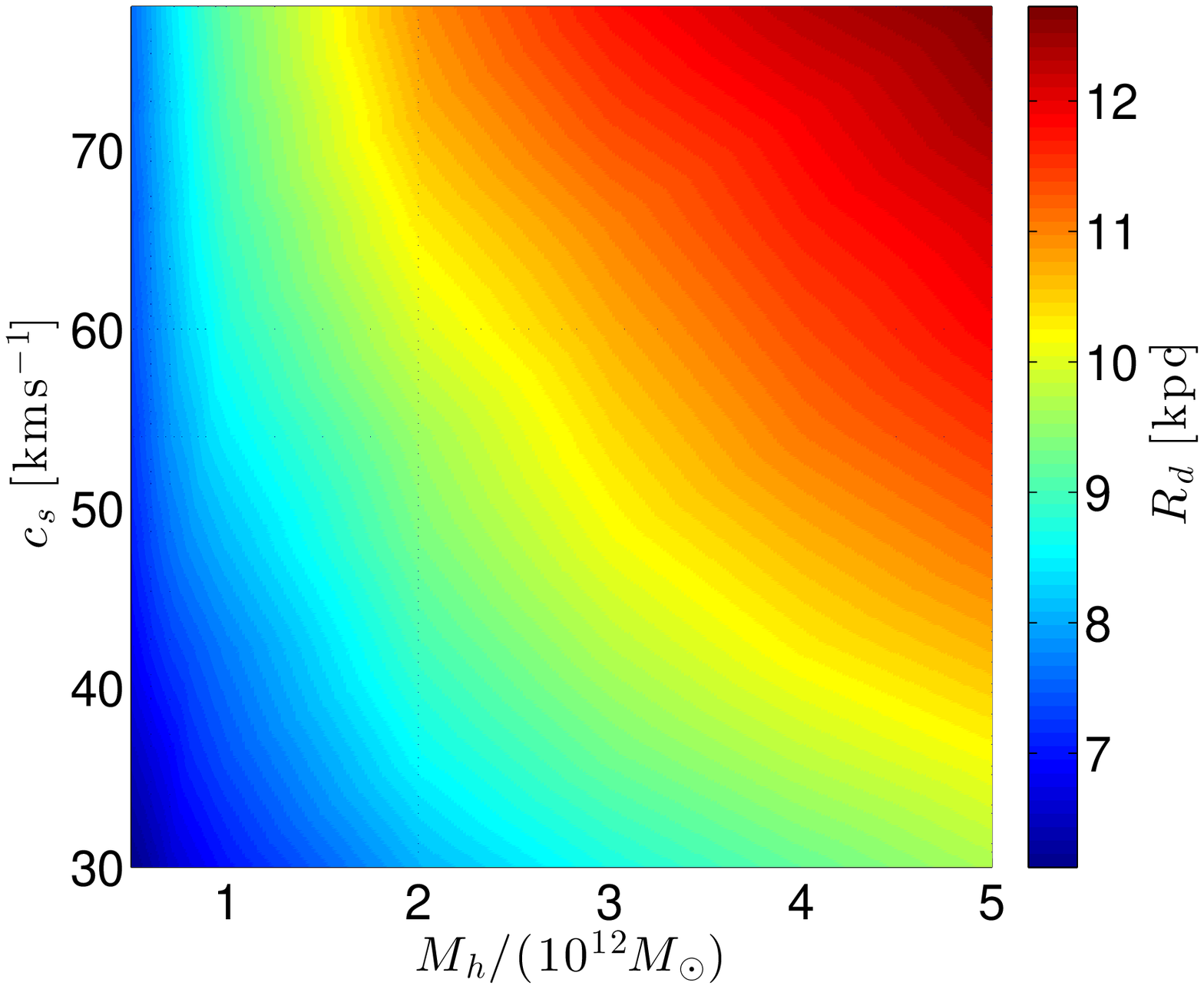}
\includegraphics[width=0.975\columnwidth]{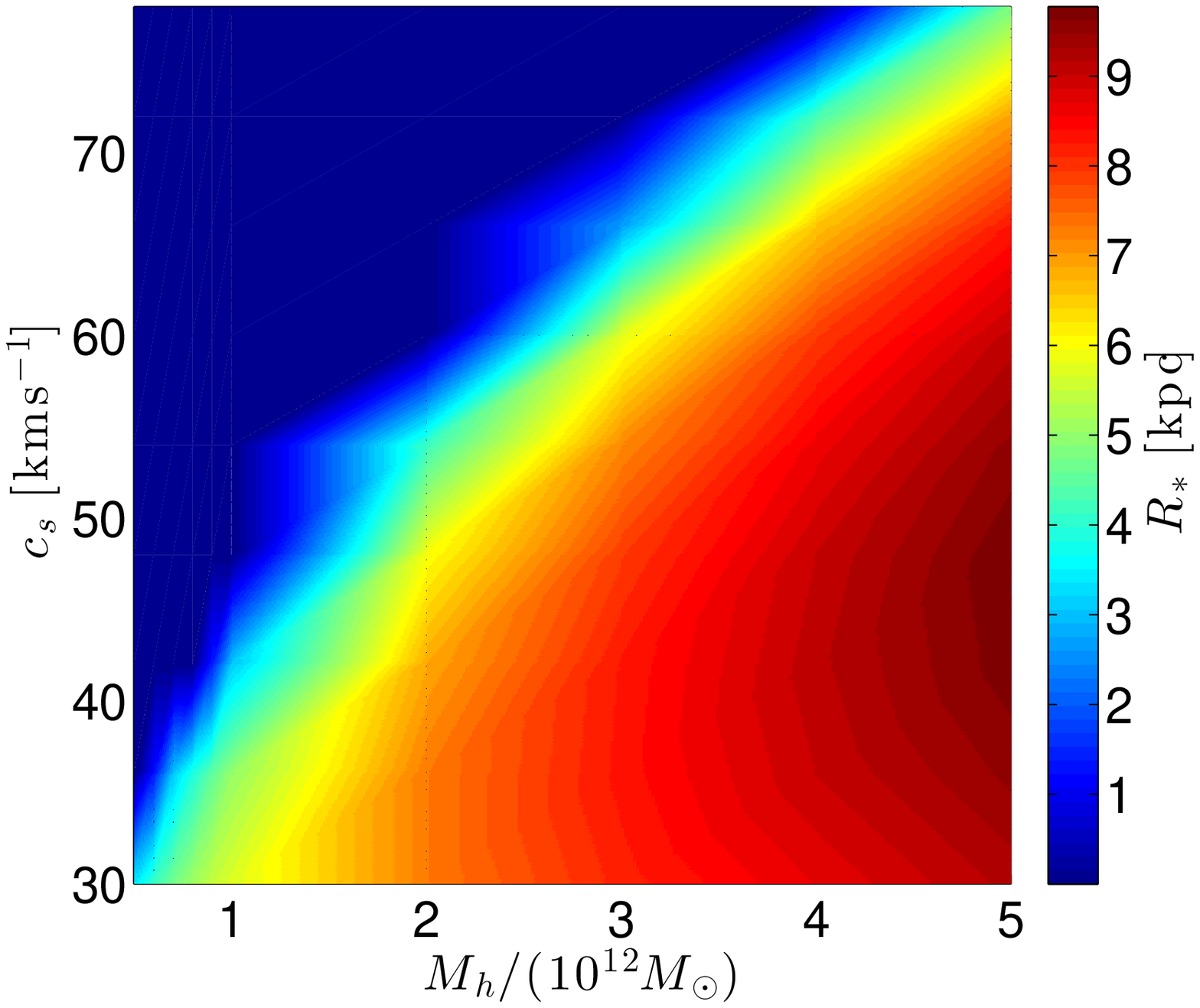}
\includegraphics[width=\columnwidth]{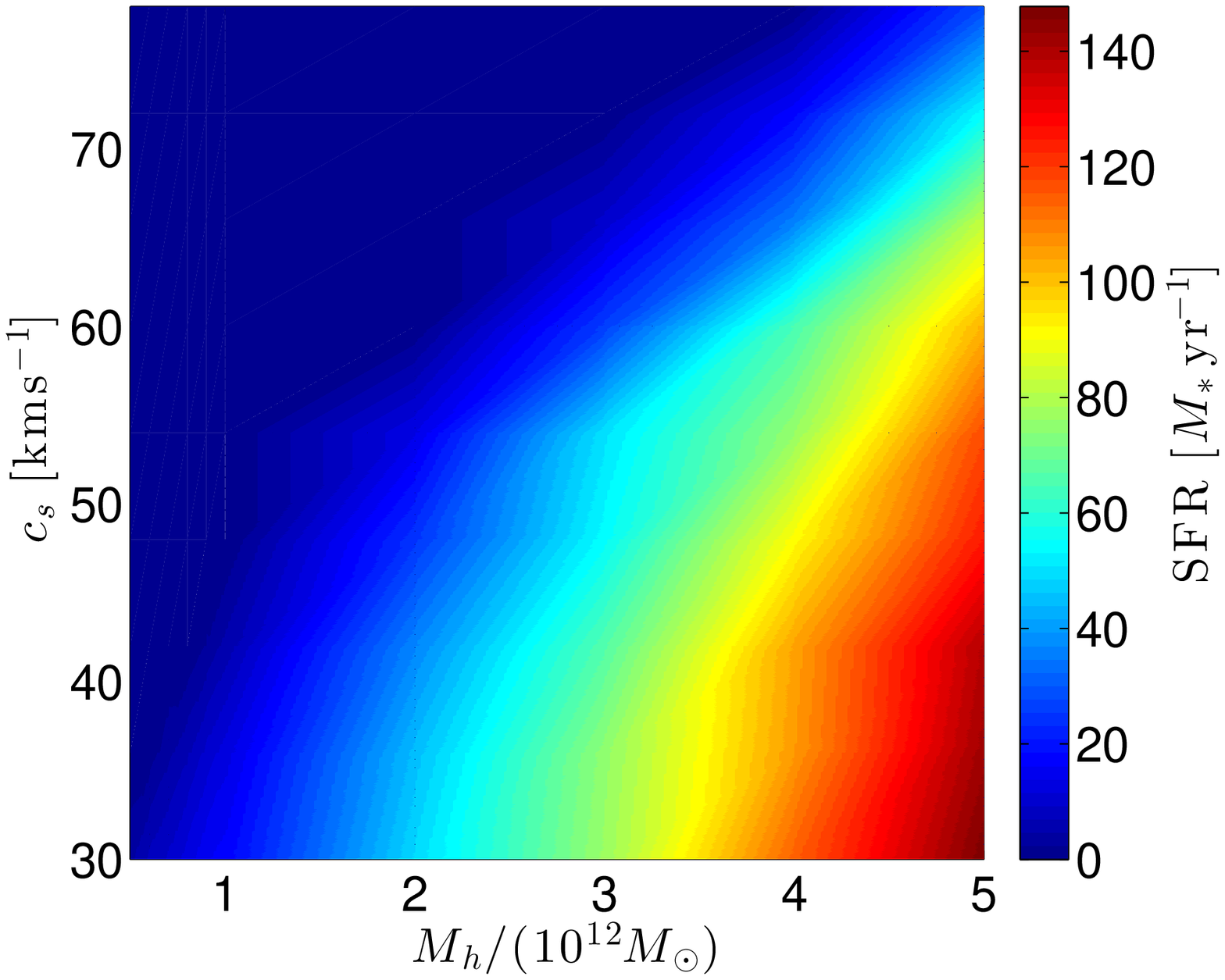}
\includegraphics[width=\columnwidth]{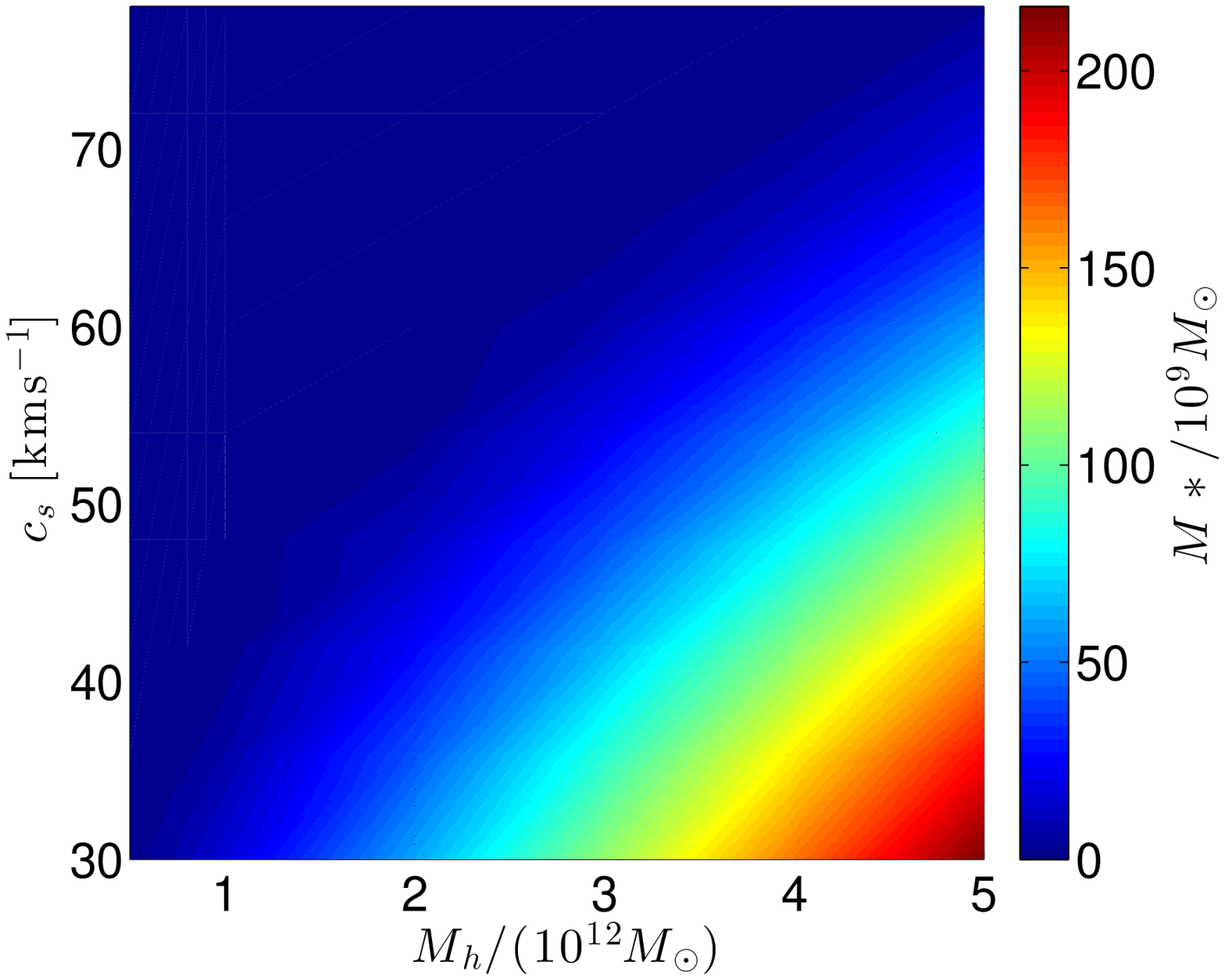}
\caption{Disk size (\emph{top left}), stellar disk size (\emph{top right}), SFR (\emph{bottom left}) and the stellar mass (\emph{top left}) as predicted by our model at $z=2$ as a function of $c_s$ and $M_h$, for $c_s=18(1+z)\,{\rm km}\,{\rm s}^{-1}$ and $\lambda = 0.035$.}
\label{fig:changeM}
\end{figure*} 
So far we have only considered the evolution of disks within a DM halo of mass $5\times10^{12}M_\odot$ at $z=2$, and examined how perturbations to  model parameters affect the evolution of those disks. We now wish to consider less massive halos. These imply smaller baryonic accretion rates, and hence less massive baryonic disks. This affects the disk size and SFR, and consequently also the gas fractions in the disks. Reduced halo masses lead to smaller disks as less gas is accreted, but as shown in Fig.~\ref{fig:changeM} (\emph{top left}), however,  the dependence of the disk size on the halo mass weakens as $c_s$ decreases. 

Since a less massive halo means reduced baryonic accretion, and hence lighter gaseous disks, it is also to be expected that a smaller $M_h$ implies smaller SFR, for a given $c_s$. As we have already shown, as $c_s$ increases, the characteristic gas surface density has to become larger to sustain star formation. Therefore, for each $c_s$ there is the corresponding $M_h$ below which star formation is no longer possible, since $Q>1$ at all times for all radii. This forms a star-formation threshold, which is visible in Fig.~\ref{fig:changeM} (\emph{bottom left}) of the SFR as a function of $c_s$ and $M_h$ at $z=2$. This naturally points out a validity range for our model.

Examining the stellar disk size as a function of $M_h$ and $c_s$, as in Fig.~\ref{fig:changeM} (\emph{top right}), we find that the behaviour we found in the analysis of Fig.~\ref{fig:cs_l_first} (\emph{middle right}), of our fiducial halo, is modified for less massive halos. For halos of mass $\lesssim3\times10^{12}M_\odot$ the stellar disk size is monotonically decreasing as $c_s$ increases. This implies that for the less massive halos the fact that the total size of the gaseous disk increases as $c_s$ increases is subdominant to the $Q$-barrier to star formation due to the increase of the gas sound speed. 

Finally, in Fig.~\ref{fig:changeM} (\emph{bottom right}) we plot the total stellar mass in the disk as a function of $M_h$ and $c_s$ at $z=2$. We find that even though the total stellar mass increases monotonically with $M_h$, as might be expected, it does not follow the general behavior of the stellar disk size discussed above, which means that maximising the star forming region does not guarantee a higher total SFR through that region.

\section{Summary}
\label{sec:discussion}

In this paper we have presented a ``radially-resolved-equilibrium-model" for the growth of baryonic disks in cosmologically accreting massive halos.  We have solved the time-dependent equations that govern the radially dependent gaseous and stellar contents of disks, as well as their star-formation rates, inflows and outflows from and to the inter- and circum-galactic medium, and inward radial gas flows within the disks, while imposing mass and angular momentum conservation. We have focused on the evolution up to  redshifts $z\sim 2$. The equilibrium model for galaxy evolution has been widely employed to account for the interrelated observed parameters of the (blue) main-sequence \citep{Bouche,Dave-2,Lilly,Dekel}. In this paper we have generalised the analytic model to account for the global disk structures --- their gaseous and stellar surface densities, radially resolved SFR --- and study the impact of the variation of various parameters on this structure. We have shown that:
\begin{itemize}
\item  The dynamical time scales required for both components of  baryonic disks to reach their equilibrium configuration are much shorter than the those of cosmological  variations in the DM halo mass and profile. This  feature guarantees that the disk structures and star formation rates remain close to equilibrium during the cosmological evolution.
\item  Radial (in plane) gas inflows together with mass conservation naturally give rise to (quasi-)exponential disks over many length scales. In our models the disk sizes, surface densities, and star formation rates are self-consistently computed assuming global conservation of the halo specific angular momentum. 
\item The stellar disk size is determined by the SFR, which is self consistently calculated according to the condition Toomre $Q<1$. For a roughly flat rotation curve, this condition is not satisfied when the surface density becomes too small, as in the outer parts of the gaseous disk, which naturally  leads to stellar disks that are more compact than the corresponding gaseous disks.
\item Focusing on cosmological evolution up to $z=2$, and on massive, $5\times 10^{12}M_\odot$ halos (at $z=2$), our model reproduces the star formation main sequence, ${\rm SFR}(M_*)$. In particular, our fiducial model predicts SFRs of $\sim 100 \, M_\odot \, {\rm yr}^{-1}$, stellar masses $\sim 9\times 10^{10}\, M_\odot$, gas contents $\sim 10^{11}\, M_\odot$, half mass sizes of 4.5 and 5.8 kpc for the stars and gas, and characteristic surface densities $500, 400\, M_\odot\, {\rm pc}^{-2}$.
\end{itemize}
Our analytic models may be used to study the interrelationships between spin parameter, halo mass, and redshift, on the disk gas and stellar surface densities and radially dependent star-formation rates.

\section*{Acknowledgments}
We thank Andreas Burkert, Reinhard Genzel and Linda Tacconi for many helpful discussions. We thank the referee for very helpful comments that improved our presentation. This work was supported by the DFG via German-Israel Project Cooperation grant STE1869/1-1/GE625/15-1. We also acknowledge support by the Raymond and Beverly Sackler Tel Aviv University - Harvard/ITC Astronomy Program, and Israel Science Foundation I-CORE grant 1829/12.

\appendix
\section{Steady state solution of $\alpha\neq0$}
\label{sec:SSSAN0}
For the velocity case (\textit{a}), the steady state solution of the $\alpha\neq0$ (that is, the star forming) scenario may be obtained analytically. To do that, we first rewrite the differential equation, as a classical non-homogeneous first order ordinary differential equation of the form $y' + p(x) y = q(x)$:
\begin{equation}
y' + \left( \frac{1}{x} - \alpha  \right) y = -1 ~~~.
\end{equation}
Identifying the coefficients $p(x),q(x)$ we can immediately find the solution to this equation, with the constant of integration chosen so that $y(1)=0$, to yield
\begin{equation}\label{eq:yOfAlpha}
y(x) = \frac{1-(1+\alpha)\exp\{\alpha x-\alpha\} +\alpha x }{\alpha^2 x} ~~~.
\end{equation}
It can be easily verified that for $\alpha\ll1$
\begin{equation}
y(x)=\frac{1-x^2}{2x} + \mathcal{O}\left(\alpha\right) ~~~,
\end{equation}
which converges continuously to the no-star formation case. Another notable fact is that having $\alpha\neq 0$ does not change the functional form of the divergence of $y$ towards the center of the disk, where the surface density satisfies 
\begin{equation}
y\approx f(\alpha)\frac{1}{2x}   \quad\quad \left( x \ll 1 \right) ~~~,
\end{equation}
where $f$ can be read off Eqn.~(\ref{eq:yOfAlpha}) and satisfies $ \forall \alpha >0:~f(\alpha)>0$.  Finally, we note that the same is true for case (\textit{b}): the dimensionless surface density may be calculated analytically, and shown to converge continuously to the $\alpha=0$ scenario. 

\section{Solving for $M_h(z)$}
\label{sec:MhOfZ}
Let us write Eqn.~(\ref{eq:Dekel}) in a more general way
\begin{equation}\label{eq:DekelGeneral}
\dot M _h = \Gamma M_h^b  \left( 1+z\right) ^c ~~~. 
\end{equation}
We note that 
\begin{equation}
\dot M _h = a\frac{{\rm d} M_h}{{\rm d}z} \frac{{\rm d}z}{{\rm d}a}\left(\frac{1}{a}\frac{{\rm d}a}{{\rm d}t}\right) ~~~,
\end{equation}
where $a$ is the cosmological scale factor and $z,t$ are the cosmological redshift and time as usual. Therefore Eqn.~(\ref{eq:DekelGeneral}) can be written as
\begin{equation}
\frac{{\rm d} M_h }{M_h^{-b}}= - \Gamma  (1+z)^{c-1} \frac{{\rm d}z}{H(z)} ~~~,
\end{equation}
where $H(z)$ is the Hubble constant at $z$. This equation can be solved numerically, as we do for the quantitative analysis in this paper. It is illustrative though to assume a matter dominated universe ($\Omega_m=1$), so that $H(z)=H_0(1+z)^{3/2}$, where $H_0$ is the present day Hubble constant, and solve this differential equation analytically. In a matter dominated universe then
\begin{equation}
M_h =\left\{ \frac{\Gamma}{H_0}\frac{2-2b}{3-2c} \left[(1+z)^{c-3/2} \right|_{z_0}^{z} + m_0^{1-b}\right\}^{1/(1-b)} ~~~,
\end{equation}
where $m_0 = M_h(z_0)$.

\label{lastpage}

\end{document}